\documentclass[11pt,reqno]{amsart}      
\usepackage{amssymb,amsthm}             
\numberwithin{equation}{section}        

\newcommand{\mnote}[1]{}     

\newcommand{\Vince}{{\tt *VM*}}
\newcommand{\Lars}{{\tt *LA*}}
\renewcommand{\Re}{{\mathbb R}}         
\newcommand{\la}{\langle}               
\newcommand{\ra}{\rangle}               
\newcommand{\half}{\frac{1}{2}}         
\newcommand{\third}{\frac{1}{3}}        

\newcommand{\Flat}{\mathbb F}		
\newcommand{\tens}{\otimes}             
\newcommand{\SO}{\text{\rm SO}}         

\newcommand{\MSlice}{\mathcal S}        
\newcommand{\CSlice}{\Sigma}		
\newcommand{\Constr}{{\mathcal C}}      
\newcommand{\Hess}{{\text{\rm Hess}}}   
\newcommand{\Weyltens}{C}                       
\newcommand{\Ric}{\text{\rm Ric}}       
\newcommand{\Scal}{\text{\rm Scal}}     
\newcommand{\starW}{{^*W}}              
\newcommand{\Wstar}{{W^*}}              
\newcommand{\hpi}{\hat \pi}             
\newcommand{\hk}{\hat k}                
\newcommand{\BR}{Q}                     
\newcommand{\EBR}{{\mathcal Q}}         
\newcommand{\Energy}{\mathcal E}        
\newcommand{\GG}{\mathcal G}            

\newcommand{\aM}{{\bar M}}              
\newcommand{\anabla}{{\bar \nabla}}     
\newcommand{\ame}{{\bar g}}             
\newcommand{\agamma}{{\bar \gamma}}     
\newcommand{\ah}{{\bar h}}              
\newcommand{\aR}{{\bar R}}              

\newcommand{\tE}{\tilde E}
\newcommand{\tB}{\tilde B}

\newcommand{\tg}{\tilde g}
\newcommand{\tk}{\tilde k}
\newcommand{\tLapse}{\tilde \Lapse}
\newcommand{\tT}{\tilde{T}}             
\newcommand{\tEBR}{\widetilde{\EBR}}    

\newcommand{\logtime}{\sigma}		
\newcommand{\Diff}{{\mathcal D}}        
\newcommand{\DD}{\mathbf D}			
\newcommand{\MM}{{\mathcal M}}          
\newcommand{\UU}{{\mathcal U}}          

\newcommand{\TT}{\text{\sc TT}}         
\newcommand{\Vol}{\text{\rm Vol}}       
\newcommand{\Eps}{\epsilon}             
\newcommand{\Lie}{{\mathcal L}}         
\newcommand{\tr}{{\text{\rm tr}}}               
\renewcommand{\div}{\text{\rm div}}         
\newcommand{\Div}{\text{\rm Div}}           
\newcommand{\curl}{\text{\rm curl}}         
\newcommand{\Lapse}{N}                  
\newcommand{\hLapse}{\hat N}            
\newcommand{\hme}{\hat g}		
\newcommand{\hnabla}{\hat \nabla}	
\newcommand{\hGamma}{\hat \Gamma}	
\newcommand{\Id}{\text{Id}}		
\newcommand{\eps}{\epsilon}		

\newcommand{\Shift}{X}                  
\newcommand{\Smallset}{\mathcal B}         

\newcommand{\length}{\text{\em (length)}}

\theoremstyle{plain}
\newtheorem{thm}{Theorem}[section]
\newtheorem{cor}[thm]{Corollary}
\newtheorem{lemma}[thm]{Lemma}

\newtheorem{definition}[thm]{Definition}
\newtheorem{prop}[thm]{Proposition}
\newtheorem{remark}{Remark}[section]

\title{Future complete vacuum spacetimes}
\author[L. Andersson]{Lars Andersson$^1$}
\thanks{$^\dagger$Supported in part by the Swedish Natural
Sciences Research Council (SNSRC),  contract no.  R-RA 4873-307 and NSF,
contract no. DMS 0104402.}
\address{Department of Mathematics\\
University of Miami\\
Coral Gables, FL 33124\\
USA}
\email{larsa\char'100math.miami.edu}

\author[V. Moncrief]{Vincent Moncrief$^2$}
\thanks{$^2$Supported in part by the NSF, with grants PHY-9732629 and
PHY-0098084 to Yale University}
\address{Department of Physics\\
Yale University\\
P.O. Box 208120\\
New Haven, CT 06520, USA}
\email{vincent.moncrief@yale.edu}

\begin{document}

\setcounter{tocdepth}{1}
\date{March 11, 2003}


\allowdisplaybreaks[2]

\begin{abstract}
In this paper we prove a global existence theorem, in the direction of
cosmological expansion, for sufficiently small perturbations of a family of
spatially compact variants of the $k=-1$ Friedmann--Robertson--Walker vacuum
spacetime. We use a special gauge defined by constant mean curvature slicing
and a spatial harmonic coordinate condition, 
and develop energy estimates through the
use of the Bel-Robinson energy and its higher order generalizations. In
addition to the smallness condition on the data, we need a topological
constraint on the spatial manifold to exclude the possibility of a
non--trivial moduli space of flat spacetime perturbations, since the latter
could not be controlled by curvature--based energies such as those of
Bel--Robinson type. Our results also demonstrate causal geodesic completeness
of the perturbed spacetimes (in the expanding direction) and establish
precise rates of decay towards the background solution which serves as an
attractor asymptotically. 
\end{abstract}

\maketitle


\section{Introduction}

In this paper we establish global existence and asymptotic behavior, in the
cosmologically expanding direction, for a family of spatially compact, vacuum
solutions to the 3+1 dimensional Einstein equations for sufficiently small
perturbations of certain known ``background'' solutions. The backgrounds we
consider are the spatially compactified variants of the familiar vacuum
$k=-1$ Friedmann--Robertson--Walker (FRW) solution, which exist on any
4--manifold $\aM$ of the form $(0,\infty) \times M$, where $M$ is a compact
hyperbolic 3--manifold (i.e., a manifold admitting a Riemannian metric with
constant negative sectional curvature). 

Let $\gamma$ be the standard hyperbolic metric with sectional curvature $-1$
on $M$. Then $(\aM, \agamma)$ given by 
\begin{align*}
\aM &= (0,\infty) \times M \\
\agamma &= - d\rho\tens d\rho + \rho^2 \gamma
\end{align*}
is a flat spacetime, locally isometric to the $k=-1$ vacuum FRW model, which
we shall call a hyperbolic cone spacetime. It has a big bang singularity as
$\rho \searrow 0$ but expands to infinite volume as $\rho \nearrow
\infty$. The vector field $\rho \frac{\partial}{\partial \rho}$ is a timelike
homothetic Killing field on $(\aM, \agamma)$ so that these backgrounds are
continuously self--similar. We shall be considering sufficiently small
perturbations of such hyperbolic cone spacetimes to the future of an
arbitrary $\rho = $constant Cauchy surface under the additional topological
restriction that $(\aM, \agamma)$ be ``rigid'' in a sense that we shall
define more fully below. The rigidity assumption serves to eliminate the
possibility of making non--trivial but still flat perturbations of the chosen
backgrounds. 

Our main result treats the vacuum Einstein equations on $\aM$ and proves
global existence in the expanding direction for initial data sufficiently
close to data for $(\aM, \agamma)$. More precisely, we show that the maximal
globally hyperbolic future vacuum development $(\aM, \ame)$  of such data is
causally geodesically complete and globally foliated by constant mean
curvature (CMC) hypersurfaces in the expanding direction. We further show
that the metric $\ame$ decays asymptotically to $\agamma$ at a well--defined
rate (that correctly predicted by linearized theory) and give the sharp rate
of decay. In this sense our result may be viewed as a nonlinear stability
result for the future evolution. We could also view it as implying nonlinear
instability for the past evolution but, since our arguments are insufficient
to treat global evolution in the past direction, we shall concentrate here on
the expanding direction. Since the formation of black holes would be expected
to violate geodesic completeness towards the future, we can also interpret
our smallness condition in the data as sufficient to exclude the formation of
black holes. 

We work in a specific gauge defined by constant mean curvature slicing and a
spatial harmonic coordinate condition which serves to kill off certain second
order terms in the spatial Ricci tensor, reducing it to a nonlinear elliptic
operator on the metric. This in turn effectively reduces the evolution
equations for the spatial metric to nonlinear wave equations wherein,
however, the lapse function and shift vector field are determined by an
associated set of (linear) elliptic equations. Local existence and
well--posedness for the Einstein equations in this gauge was established in
\cite{andersson:moncrief:local} along with a continuation principle which
provides the needed criterion for proving global existence. 

The main tool we employ for our global existence proof is an energy argument
based on the Bel--Robinson energy and its higher order generalization, which
we define. The Bel--Robinson energy for a vacuum spacetime is basically an
$L^2$--norm of spacetime curvature on a given Cauchy hypersurface, and its
higher order generalization incorporates the $L^2$--norm of the spatial
gradient of this same curvature. One of the key steps in our proof will be to
show that, in our chosen gauge, this generalized Bel--Robinson energy bounds
an $H^3\times H^2$ norm of the perturbed first and second fundamental forms
of a CMC slice in the spacetime $(\aM, \ame)$. 

Nontrivial spacetime perturbations which preserve flatness are invisible to
such purely curvature based energies, and this is the reason we have been
forced to impose an additional rigidity condition upon the hyperbolic
manifolds that we consider. Already by Mostow rigidity one cannot perturb the
flat metric $\agamma = - d\rho \tens d\rho + \rho^2 \gamma$ to another flat
one by simply deforming the hyperbolic metric $\gamma$ on $M$, but there can
be more subtle ways of deforming $\agamma$ on $\aM$ that preserve
flatness. These arise whenever $(M,\gamma)$ admits so--called nontrivial
traceless Codazzi tensors. Our rigidity requirement is that $(M,\gamma)$ be
such as to exclude such tensors---a condition which is known to be satisfied
for a non--empty set of hyperbolic manifolds. 

The Bel--Robinson energy is of course not a conserved quantity but, together
with its higher order generalization, can actually be shown to decay in the
expanding direction for sufficiently small perturbations of a hyperbolic cone
spacetime. The main source of this decay is the overall expansion of the
universe which leads to an omnipresent term of good sign, proportional to the
energy itself, in the time derivative of this energy. A corresponding result
holds for the generalized energy. The remaining terms in the time derivative
in general have no clear sign but fortunately can be bounded by a power
greater than unity of the generalized energy itself. When the initial value
of the generalized energy is sufficiently small this implies decay to the
future at an asymptotically well--defined rate and leads to our main
result. 

While we shall not pursue this issue here, there seems to be a
straightforward way to remove the rigidity constraint and thereby deal with
arbitrary hyperbolic $M$. This involves supplementing the Bel--Robinson
energies considered here by another non--curvature--based energy called the reduced
Hamiltonian. As discussed in \cite{fischer:moncrief:hamred} this quantity is
always monotonically decaying towards the future (even for large data) but
bounds at most the rather weak $H^1\times L^2$ norm of the CMC Cauchy
data. However this should more than suffice to control the finite dimensional
space of moduli parameters which arises in the case of non--rigid $M$ but is
invisible to the Bel--Robinson energies. 

Apart from general Lorentzian geometry results such as singularity theorems
and conclusions drawn from the study of explicit solutions, very little is
known about the global properties of generic 3+1 dimensional Einstein
spacetimes, with or without matter, and present PDE technology is far from
being applicable to the study of such global questions, except in the case of
small data. 

In \cite{christo:klain} Christodoulou and Klainerman proved the
nonlinear stability of 3+1 dimensional Minkowski space, i.e., a small data
global existence result together with precise statements about the asymptotic
decay of the metric to the Minkowski metric. This proof was based on a
bootstrap argument using decay estimates for suitably defined Bel--Robinson
energies. A central element in the proof was the construction of approximate
Killing and conformal Killing fields, which were then used in a way which is
analogous to the way in which true Killing and conformal Killing fields of
Minkowski space are used in the proof of the Klainerman Sobolev inequalities
for solutions of the wave equation on Minkowski space. 

In still earlier work \cite{friedrich:complete} Friedrich had proven global existence to the
future of a Cauchy surface for the development of data sufficiently close to
that of a hyperboloid in Minkowski space, with asymptotic behavior
compatible with a regular conformal compactification in the sense of
Penrose. This result used the fact that the conformal compactification of
such spacetimes has a regular null boundary (Scri) and exploited Cauchy
stability for a conformally regular first order symmetric hyperbolic system
of field equations deduced from the Einstein equations. Roughly speaking,
local existence for the conformally regular system can correspond, for
sufficiently small data, to global existence for the conformally related,
physical spacetime. 

Our argument is close in spirit to that of Christodoulou and Klainerman but
is much simpler than theirs by virture of the universal energy decay
described above. The source of this decay can easily be seen in linear
perturbation theory by exploiting the fact that $\rho
\frac{\partial}{\partial \rho}$ is a timelike homothetic Killing field in the
background. One readily constructs from this an exactly conserved quantity
for the linearized equations which differs from the (linearized analogue of
the) Bel--Robinson energy we consider by a multiplicative factor in the time
variable $\rho$. This gives immediately the specific decay rate predicted by
linearized theory and our arguments ultimately show that this is the precise
decay rate asymptotically realized by solutions to the (small data) nonlinear
problem. 

Our arguments are also similar in spirit to those of \cite{ChB:moncrief:U(1)} 
in which
Choquet--Bruhat and Moncrief treat perturbations of certain
$\UU(1)$--symmetric vacuum spacetimes on $\Re\times \Sigma \times S^1$, where
$\Sigma$ is a higher genus surface and in which the $\UU(1)$ (Killing)
symmetry is imposed along the circular fibers of the product bundle
$\Re\times \Sigma \times S^1 \to \Re\times \Sigma$. Their results also use
energy arguments which exploit the universal expansion to obtain decay for
small data. A significant generalization of that work is presented in the
article by Choquet--Bruhat in the present volume, wherein she removes the
restriction to ``polarized'' solutions adopted in the earlier work. For the
case of linearized perturbations, Fischer and Moncrief 
\cite{fisher:moncrief:n+1} have
analyzed the stability of higher dimensional analogues of the hyperbolic cone
spacetimes described above wherein the hyperbolic metric $\gamma$ is replaced
by an arbitrary Einstein metric with negative Einstein constant. These of
course include the higher dimensional hyperbolic metrics but in fact comprise
a much larger set. It now seems likely that the nonlinear stability problem
for these spacetimes can be handled by a combination of the methods employed
herein and in the article by Choquet--Bruhat. 

We now give a more precise description of our main results. Let $g$ be a
Riemannian metric on $M$ and let $k$ be a symmetric covariant 2--tensor on
$M$. We call $(M,g,k)$ a vacuum data set for the Einstein equations if
$(g,k)$ satisfy the vacuum constraint equations, reviewed in section
\ref{sec:vacEin} below. Given such a vacuum data set there is a unique
maximal Cauchy development $(\aM, \ame)$ of $(M,g,k)$ which contains the
latter as an embedded Cauchy hypersurface. Our results concern the structure
of $(\aM, \ame)$, especially to the future of the Cauchy hypersurface, for
$(g,k)$ sufficiently close to the data corresponding to a rigid 
\mnote{added rigid, changed to ``hyperbolic cone''}
hyperbolic cone spacetime $(\aM,\agamma)$. We show in this case that, in the 
expanding direction, $(\aM, \ame)$ is globally foliated by hypersurfaces of
constant mean curvature and that $(\aM, \ame)$ is causally geodesically
complete in this (future) direction. In particular, $(\aM, \ame)$ is
inextendible in the expanding direction and thus our results support the
strong cosmic censorship hypothesis. 

Our main result is summarized as follows. 

\begin{thm}\label{thm:main-intro}
Let $(M,\gamma)$ be a compact hyperbolic 3--manifold and assume that
$(M,\gamma)$ is rigid (i.e., admits no nontrivial traceless Codazzi
tensors). Assume that $(M, g^0, k^0)$ is a CMC vacuum data set with $(g^0,
k^0) \in H^s\times H^{s-1}$, $s \geq 3$, having $t_0 = \tr_{g^0} k^0 =
constant < 0$. Then there is an $\eps > 0$ so that if 
$$
||\frac{t_0^2}{9} g^0 - \gamma ||_{H^3} + ||\frac{t_0}{3} k^0 - \gamma ||_{H^2}
  < \eps
$$
then 
\begin{enumerate}
\item The maximal Cauchy development $(\aM, \ame)$ of the vacuum data set
$(M, g^0, k^0)$ has a global CMC foliation in the expanding direction (to
the future of $M_{t_0}$ in CMC time $t = \tr_g k$). 
\item $(\aM, \ame)$ is future causally geodesically complete.
\end{enumerate}
\end{thm}
\begin{remark}
\begin{enumerate}
\item Under our conventions, c.f. section \ref{sec:prel}, the standard
hyperboloid $\{ \la x,x\ra = -1 \}$ in $I^+ (\{0\}) \subset \Re^{3,1}$ has
mean curvature $-3$ and $\Vol(M,g)$ increases as $t \nearrow 0$. 
\item $(\frac{t^2}{9} g, \frac{|t|}{3} k)$ are rescaled Cauchy data that
reduce to $(\gamma, - \gamma)$ for the background solution. Our energy
arguments show that the rescaled data approach their background values at a
well--defined asymptotic rate as $t = \tr_g k \nearrow 0$. 
\item By exploiting the scaling with respect to $t$ at $t_0$ one can satisfy
the smallness condition for initial data $(g^0, k^0)$ corresponding to
arbitrarily large initial spacetime curvature. In this sense one can choose the
initial hypersurface to be ``close to the singularity''. 
\end{enumerate}
\end{remark}

In outline our paper proceeds as follows. Some preliminaries and a discussion
of the Einstein equations in our chosen gauge including a review of the local
existence theorem proven in \cite{andersson:moncrief:local}, are given in
sections \ref{sec:prel} and \ref{sec:vacEin}.  Sections
\ref{sec:lobell}--\ref{sec:flatsp} discuss the background spacetimes, the
constraint set for the perturbed spacetimes and the rigidity condition needed
to exclude the occurrence of a moduli space of flat perturbations. Section 3
introduces Weyl fields in the spirit of Christodoulou and Klainerman and
presents the field equations they satisfy when Einstein's equations are
imposed. 
Section \ref{sec:BRenergy} discusses the Bel--Robinson energy and its higher
order generalization and computes the time derivative of these quantities in
the chosen gauge. Section \ref{sec:BRvac} describes the scale--free variants
of these energies that are used in our estimates and section \ref{sec:BRhess}
gives the calculation which shows how these energies actually bound Sobolev
norms of the perturbed data in the rigid case. Sections \ref{sec:estimates}
and \ref{sec:BRdiff} discuss estimates and the differential inequalities
satisfied by our rescaled Bel--Robinson energies. The global existence proof
is completed in section \ref{sec:global} and section \ref{sec:complete}
establishes causal geodesic completeness. A number of useful definitions and
identities are collected in the appendix.

\section{Preliminaries\label{sec:prel}}
Let $\aM$ be a spacetime, i.e 
an n+1 dimensional manifold with Lorentz metric $\ame$ of
signature $-$$+\dots+$ and covariant derivative $\anabla$. We denote by 
$\la \cdot , \cdot \ra$ the scalar product defined by $\ame$ on $T\aM$. 
Let $M \subset \aM$ be a spacelike hypersurface 
with timelike normal $T$, $\la T, T \ra = -1$, and let 
$t$ be a time function on a neighborhood of $M$.
Then we can introduce
local coordinates $(t,x^i, i=1,\dots,n)$ on $\aM$ so that $x^i$ are coordinates
on the level sets $M_t$ of $t$. We will often drop the subscript $t$ on $M_t$
and associated fields. 

Let $\partial_t =\partial/\partial t$ be the coordinate vector field 
corresponding to $t$.  
The lapse function $\Lapse$ and shift vectorfield $\Shift$ of the foliation
$\{M_t\}$ are defined by
$
\partial_t = \Lapse T + \Shift .
$
Assume $T$ is future directed so that $N > 0$. 
The space--time metric $\ame$ takes the form
\begin{equation}\label{eq:n+1metr}
\ame = - \Lapse^2 dt \tens dt 
+ g_{ij} (dx^i + \Shift^i dt )\tens (dx^j + \Shift^j dt) .
\end{equation}
Let $\{e_i\}_{i=1,\dots,n}$ be a Fermi-propagated
orthonormal frame tangent to  $M_t$,
i.e. 
$
\la \anabla_T e_i , e_j \ra = 0,\  \forall
i,j , 
$
with dual frame $\{e^i\}_{i=1}^n$.
If one drops the assumption that the frame is Fermi propagated,
then in general $\anabla_{T} e_i = \anabla^{//}_{T} e_i + 
(\Lapse^{-1} \nabla_i \Lapse) T $, where  $\anabla^{//}_T e_i$ denotes the
tangential part of $\anabla_T e_i$. With $e_0 = T$, 
$\{e_{\mu}\}_{\mu=0}^n$ is an ON frame on $\aM$, adapted to the
foliation $\{M_t\}$. We will use the convention that lower 
case latin indices run over over $1,\dots,n$, while greek indices run over 
$0,\dots, n$. Our conventions for curvature as well as some useful identities
are given in Appendix \ref{sec:basic}.
The index
$T$ in a tensor expression denotes contraction with $T$,  for example
$\anabla_{T} A_\alpha = T^\beta \anabla_\beta A_\alpha$. 


The second fundamental form $k_{ij}$ of $M_t$ is given by
$
k_{ij} = -\half (\Lie_T \ame)_{ij} .
$
In terms of the Fermi-propagated frame $\{e_i\}$ we have the following 
relations 
between $\Lapse, T$ and $k_{ij}$. 
\begin{subequations}\label{eq:framerel}
\begin{align}
\anabla_i e_j &= \nabla_i e_j - k_{ij} T , &
\anabla_i T &= - k_{ij} e_j , \\
\anabla_{T} e_i &= (\Lapse^{-1} \nabla_i \Lapse) T , &
\anabla_T T &= (\Lapse^{-1} \nabla_i \Lapse) e_i .
\end{align}
\end{subequations}
In
computations we frequently make use of equations (\ref{eq:framerel}) to do an
n+1 split, for example $\anabla_i A_j  = \nabla_i A_j + k_{ij} A_{T}$. 

\subsection{The vacuum Einstein equations} \label{sec:vacEin}
The vacuum Einstein equations 
\begin{equation}\label{eq:vacuum}
\aR_{\alpha\beta}=0 ,
\end{equation}
consist after a n+1 split of the 
constraint equations
\begin{subequations}\label{eq:constraint}
\begin{align}
R - |k|^2 + (\tr k)^2 &= 0 , \label{eq:constraint-ham} \\
\nabla_i \tr k - \nabla^j k_{ij}   &= 0 , \label{eq:constraint-mom}
\end{align}
\end{subequations}
and the evolution equations
\begin{subequations}\label{eq:evolution}
\begin{align}
\Lie_{\partial_t} g_{ij} &= -2 \Lapse k_{ij} + \Lie_{\Shift} g_{ij} , 
\label{eq:evolution-g}\\
\Lie_{\partial_t} k_{ij} &= - \nabla_i \nabla_j \Lapse + \Lapse(R_{ij} + \tr k
k_{ij} - 2 k_{im}k^m_{\ j} ) + \Lie_{\Shift} k_{ij} .\label{eq:evolution-k}
\end{align}
\end{subequations}
We will call a solution $(g^0, k^0)$ to the Einstein vacuum constraint 
equations on $M$, a {\bf vacuum data set}. A curve $t \mapsto
(g,k,\Lapse,\Shift)$ solving the  Einstein vacuum 
evolution and constraint equations corresponds to a vacuum space--time metric 
$\ame$ via (\ref{eq:n+1metr}). A vacuum 
space--time $(\aM, \ame)$ with an isometric
imbedding of a vacuum data set $(g,k)$ on $M$ is said to be a {\bf vacuum 
extension} of $(g,k)$. 

Let $\hme$ be a fixed $C^{\infty}$ Riemann metric on $M$ with Levi--Civita 
covariant derivative $\hnabla$ and Christoffel symbol $\hGamma^k_{ij}$.  
Define
the vector field $V^k$ by
\begin{equation}\label{eq:Vdef}
V^k = g^{ij} e^k ( \nabla_i e_j - \hnabla_i e_j) . 
\end{equation}
In terms of a coordinate frame, $V^k = g^{ij} (\Gamma^k_{ij} -
\hGamma^k_{ij})$. The identity map $\Id: (M,g) \to (M,
\hme)$ is harmonic exactly when $V^k = 0$, see
\cite{andersson:moncrief:local} for discussion. 

A vacuum data set $(g,k)$ is in {\bf CMCSH gauge} with respect to $\hme$ if
\begin{subequations}\label{eq:gauge}
\begin{align}
\tr_g k &= t  && \text{\rm (Constant Mean Curvature),} \label{eq:gauge-CMC}\\
V^k  &= 0  &&\text{\rm (Spatial Harmonic
coordinates),} \label{eq:gauge-SH}
\end{align}
\end{subequations}
Given a space--time $(\aM, \ame)$, a foliation 
$\{M_t, \ t \in (T_-, T_+)\}$ in $(\aM, \ame)$ is called 
a CMC foliation if $\nabla \tr k = 0$ for all $t \in (T_-,T_+)$. In this case, we
may after a change of time parameter assume $t = \tr k$. 
If the induced data 
$(g,k)$ on $M_t$ is in CMCSH gauge for 
all $t\in (T_-, T_+)$, then $\{M_t\}$ is called a CMCSH foliation. 
The CMCSH gauge conditions imply the following elliptic equations for the
lapse and shift 
\begin{subequations}\label{eq:defineNX}
\begin{align}
- \Delta \Lapse + |k|^2 \Lapse &= 1 ,  \label{eq:defineN} \\
\Delta \Shift^i + R^i_{\ f} \Shift^f - \Lie_{\Shift} V^i 
&= (- 2 \Lapse k^{mn} + 2 \nabla^m \Shift^n )
e^i(\nabla_m e_n - \hnabla_m e_n)  
\nonumber \\
&\quad
+ 2 \nabla^m \Lapse k_m^i - \nabla^i \Lapse k_m^{\ m} ,
\label{eq:defineX}
\end{align}
\end{subequations}
where $\Delta X^i = g^{mn} \nabla_m \nabla_n X^i$.
The ellipticity constant $\Lambda[g]$ of $g$ is defined
as the least $\Lambda \geq 1$ so that
\begin{equation}\label{eq:gunif}
\Lambda^{-1} g(Y,Y) \leq \hme (Y,Y) \leq \Lambda g(Y,Y) ,
\quad \forall Y \in TM .
\end{equation}
Let $\ame$ defined in terms of $g,\Lapse,\Shift$ by
(\ref{eq:n+1metr}). Define $\Lambda[\ame]$ by
\mnote{\Vince this is not scale invariant, need definition of $||\cdot ||$}
\begin{equation}\label{eq:Lambda-ame-def}
\Lambda[\ame] = \Lambda[g] + || \Lapse ||_{L^{\infty}} 
+ || \Lapse^{-1}||_{L^{\infty}} 
+ || \Shift ||_{L^{\infty}} .
\end{equation}
Then $\ame$ is a nondegenerate Lorentz metric, as long as $\Lambda[\ame]$ is
bounded. 

We refer to \cite{andersson:moncrief:local} for the background and proof of
the following theorem and for the analysis concepts used in the present paper. 
\begin{thm}[\cite{andersson:moncrief:local}]\label{thm:AMlocal}
Assume that $M$ is of hyperbolic type with hyperbolic metric $\hme$ of 
unit negative sectional curvature. Let 
$(g^0 , k^0) \in H^s \times H^{s-1}$, $s > n/2 +1$, $s$ integer,   
\mnote{$s$ integer is not needed} 
be a vacuum data set on $M$ in CMCSH gauge with respect to
$\hme$. 
Let $t_0 = \tr k^0$. The following holds. 
\begin{enumerate}
\item \label{point:exist} {\bf Existence:}
There are $T_- < t_0 < T_+ \leq 0$ so that 
there is a vacuum extension $(\aM, \ame)$ of $(g^0, k^0)$, 
$\aM = (T_-,T_+)\times M$, 
$\ame \in H^s(\aM)$, and such that the foliation 
$\{M_t = \{t\} \times M , \ t \in (T_-,T_+)\}$,
is CMCSH. 
\item \label{point:cont} 
{\bf Continuation:} Suppose that $(T_-, T_+)$ is maximal among all
intervals satisfying point \ref{point:exist}. Then 
either $(T_-, T_+) = (-\infty, 0)$ or 
\mnote{change this to $L^1L^\infty$?}
$$
\limsup  \left ( \Lambda[\ame] + 
||D\ame||_{L^{\infty}} + || k ||_{L^{\infty}} \right )  = \infty
$$
as $t \nearrow T_+$ or as $t \searrow T_-$.
\item {\bf Cauchy stability:} 
Let $\ame$ be the space--time metric constructed from the solution
$(g,k,\Lapse,\Shift)$ to the Einstein vacuum equations in CMCSH gauge. The
map $(g^0, k^0) \to \ame$ is continuous $H^s \times H^{s-1} \to H^s((t_-,
t_+)\times M)$, for all $t_-, t_+$, satisfying $T_- < t_-<  t_+ < T_+$. 
\end{enumerate}
\end{thm}

\subsection{Hyperbolic cone space--times}\label{sec:lobell}
Let $(M,\gamma)$ be a compact manifold of hyperbolic type, 
of dimension $n \geq 2$, 
with hyperbolic metric $\gamma$ of sectional curvature $-1$.
The {\bf hyperbolic cone space--time} $(\aM, \agamma_0)$ with spatial section $M$ is
the Lorentzian cone over $(M,\gamma)$, i.e. 
$$
\aM = (0, \infty) \times M, \qquad \agamma = - d\rho^2 + \rho^2 \gamma .
$$
Let $(\aM,\agamma)$ be a hyperbolic cone spacetime of dimension $n+1$.
The family of hyperboloids $M_\rho$ 
given by $\rho = $constant has normal 
$$
T =  \partial_\rho.
$$
Here $T$ is future directed w.r.t. the time function $\rho$. 
Construct an adapted ON frame $T,e_i$ on $\aM$. 
A calculation gives
$$
k_{ij} = -\frac{1}{\rho}g_{ij} ,
$$ 
and the mean
curvature is given by $\tr k = -n/\rho$. 
The mean curvature time is defined by setting 
$t = \tr k$ and the $t$-foliation has lapse 
$$
\Lapse = - \la \partial_{t} , T \ra  = \frac{n}{t^2} .
$$
In terms of the mean curvature time we have 
\begin{equation}\label{eq:g(tau)-stand}
g(t) = \frac{n^2}{t^2} \gamma, \qquad 
k(t) = \frac{n}{t} \gamma .
\end{equation}

In the rest of this section we will consider CMCSH foliations, with the
reference metric $\hme$ chosen as $\hme = \gamma$. 

\subsection{The constraint set and the slice\label{sec:geom-slice}}
Let $M$ be a compact manifold of hyperbolic type, of dimension $n \geq 2$
with hyperbolic metric $\gamma$ of sectional curvature
$-1$. 

For $s > n/2$, let $\MM^s$ be the manifold of Riemann metrics of Sobolev
class $H^s$ on $M$. Then $\MM^s$ is a smooth Hilbert manifold 
and $\Diff^{s+1}$ acts on $\MM^s$ as a Frechet Lie group. 

\begin{lemma} \label{lem:harmonic} Let $s > n/2+1$ and fix $\tau \in \Re$,
$\tau \ne 0$. 
There is an open neighborhood $\UU^s_{\tau} \subset \MM^s$ of $\frac{n^2}{\tau^2}\gamma$, so that
for all $g\in \UU^s_{\tau}$, there is a unique $\phi \in \Diff^{s+1}(M)$, so
that $\phi : (M,g) \to (M,\gamma)$ is harmonic. 
\end{lemma} 
\begin{proof} $M$ is compact and $\gamma$ has negative sectional 
curvature. Then there is a unique
harmonic map $\phi \in H^{s+1}(M;M)$ from $(M,g)$ to $(M,\gamma)$
\cite{eells:lemaire:report1}.  
For $g$ close to $\gamma$, the implicit function theorem
shows $\phi$ is close to the identity map 
$\Id$ and hence $\phi \in \Diff^{s+1}(M)$. 
\end{proof}
Let $\UU^s_{\tau}$ be as in Lemma \ref{lem:harmonic}.
Let $\MSlice^s_{\tau} \subset \MM^s$ 
be defined by 
\begin{equation}\label{eq:slicedef}
\MSlice^s_{\tau} = \{ g \in \UU^s_{\tau}  : \Id : (M,g) \to
(M,\gamma) \text{ is harmonic} \} .
\end{equation} 
For $g \in \UU^s_{\tau}$, if $\phi$ is the harmonic map
provided by Lemma \ref{lem:harmonic}, $\phi_* g \in \MSlice^s_{\tau}$. 
By uniqueness for harmonic maps with target $(M,\gamma)$, it follows that 
$\MSlice^s_{\tau}$ is  
a local slice for the action of $\Diff^{s+1}$ on $\MM$. 
For $s > n/2$, let 
\begin{multline}\label{eq:Constrdef}
\Constr^s_{\tau} = \{ (g,k) \in H^s \times H^{s-1}, \quad \tr_g k = \tau, \\
\quad (g,k) 
\text{ solves the 
constraint equations (\ref{eq:constraint})} \} , 
\end{multline}
be the set of solutions to the vacuum Einstein constraint equations, with 
$\tr k = \tau$. As $M$ is a manifold of
hyperbolic type, $\Constr^s_{\tau}$ is a smooth Hilbert manifold, cf. 
\cite{fischer:moncrief:hamred}. 
The action of $\Diff^{s+1}$ on $\Constr^s_{\tau}$ is the lift of the action on $\MM$,
and therefore the local slice $\MSlice^s_{\tau} \subset \MM^s$ lifts to a 
local slice $\CSlice^s_{\tau}$, 
at $(\frac{n^2}{\tau^2} \gamma, k) \in \Constr^s_{\tau}$, 
$$
\CSlice^s_{\tau} =  \{ (g,k) : (g,k) \in \Constr^s_{\tau}
\text{ and } g \in \MSlice^s_{\tau} \} .
$$
The slice $\CSlice^s_{\tau}$ is a smooth Hilbert submanifold of
$\Constr^s_{\tau}$. 

A symmetric 2-tensor $h$ on $(M,g)$, which satisfies 
$\tr h = 0$, $\div h = 0$, 
is called a $\TT$--tensor (w.r.t. $g$). 
In the rest of this section, let $\DD$ denote 
the Frechet derivative in the direction $(h,p) \in T_{(\gamma, -
\gamma)}\Constr_{-n}$. It is important to
keep in mind that expressions involving $\DD$ are evaluated at $(\gamma, -
\gamma)$. 
\begin{lemma}\label{lem:TSlice} 
\begin{equation}\label{eq:TSlice}
T_{(\frac{n^2}{\tau^2}\gamma,\frac{n}{\tau}\gamma)} \CSlice_{\tau} 
= \{ (h , p) , \quad h, p \text{ $\TT$--tensors w.r.t. } \gamma \} .
\end{equation}
\end{lemma}
\begin{proof}
We give the proof assuming $\tau = -n$, the general 
case follows by scaling. 
First note
\begin{equation}\label{eq:Dtrk}
0= \DD \tr k   =  \tr_{\gamma} h + \tr_{\gamma} p .
\end{equation}
Since $k \big{|}_{(\gamma, - \gamma)} = - \gamma$, 
(\ref{eq:Dtrk}) implies $\DD|k|^2  = 0$. 
Let $H = R + (\tr k)^2 - |k|^2$, so that the 
Hamiltonian constraint (\ref{eq:constraint-ham}) is $0 = H$. By the above, 
$$
0 = \DD H  = \DD R .
$$ 
Any symmetric 2--tensor can be decomposed as 
$$
h = \psi \gamma + h_{\TT} + \Lie_Y \gamma ,
$$
where $\psi$ is a function, $h_{\TT}$ is a $\TT$--tensor and $Y$ is a
vector--field. 
As $\gamma$ is hyperbolic, $R[\gamma]$ is constant, and due to covariance of
$R$,  
$\DD R.   \Lie_{Y} \gamma = YR[\gamma]  = 0$. The 
Frechet derivative of the scalar curvature is the operator
\begin{equation*}
D R . u = - \nabla^k\nabla_k  u_i^{\ i} + \nabla^i \nabla^j u_{ij} - R_{ij}
u^{ij}  ,
\end{equation*}
which by the above gives 
\begin{align*}
\DD R . h &= \DD R . (\psi \gamma) \\
&= - (n-1) \Delta_{\gamma} \psi + n(n-1) \psi .
\end{align*}
In view of the fact that $\Delta$ is negative semidefinite, $0 = \DD R$
implies $\psi = 0$. 
Thus, 
\begin{equation} \label{eq:hform} 
h = h_{\TT} + \Lie_Y \gamma . 
\end{equation}
Let $V$ be given by (\ref{eq:Vdef}). Then with $\hme = \gamma$, 
$$
\DD V^i = \nabla^j h_j^{\ i} - \half \nabla^i \tr h ,
$$
where $\nabla,\Gamma, \tr$ are defined w.r.t. $\gamma$. 
By  definition, $V = 0$ on 
$\MSlice_{\tau}$, and therefore 
$(h,p) \in T_{(\gamma,-\gamma)} \CSlice_{\tau} $ 
implies using (\ref{eq:hform}) 
$$
\DD V = \DD V . L_Y \gamma ,
$$
which by the uniqueness of harmonic maps with target $\gamma$ implies that
$Y = 0$ and hence $h = h_{\TT}$. In particular $\tr h = 0$ and therefore 
by (\ref{eq:Dtrk}), $\tr p = 0$. 

Let $C_i = \nabla_i \tr k - \nabla^j k_{ji}$ so that the momentum constraint
(\ref{eq:constraint-mom})
is $0 = C_i$. 
By assumption, 
$\nabla_i \tr k = 0$,  which using $h = h_{\TT}$ and the momentum constraint
(\ref{eq:constraint-mom}) 
gives 
$$
0 =  \nabla^j p_{ji} ,
$$
where $\nabla$ is the covariant derivative w.r.t. $\gamma$. 
By the above,  $\tr p = 0$ so $p$ is a $\TT$--tensor w.r.t. $\gamma$. 
\end{proof} 

\subsection{Flat space--times}\label{sec:flatsp}
Let $(\aM,\agamma)$ be a hyperbolic cone space--time of dimension $n+1$, $n \geq 2$, with
spatial section $(M,\gamma)$. 
Consider a vacuum metric $\ame$ on $\aM$. Then, 
$C_{\alpha\beta\gamma\delta} = \aR_{\alpha\beta\gamma\delta}$ 
is the Weyl tensor and by the structure equations, 
\begin{align} 
C_{iT jT} &= R_{ij} - k_{im}k^m_{\ j} + k_{ij} \tr k 
\label{eq:CiTjT} \\
C_{mT ij} &= d^{\nabla} k_{mij} \label{eq:CmTij} 
\end{align}
where the covariant exterior derivative  $d^{\nabla}u$ on symmetric
2-tensors is 
$$
(d^{\nabla} u)_{ijk} = \nabla_k u_{ij} - \nabla_j u_{ik}  
$$
Let $E_{ij} = C_{iT j T}$ and $F_{mij} = C_{mT ij}$, considered
as tensors on $M$, and define the second order elliptic operator 
$A$  on symmetric 2--tensors
by 
\begin{equation}\label{eq:Adef}
A u =  \nabla^* \nabla u - n u , 
\end{equation}
so that $(Au)_{ij} = - \nabla^k \nabla_k u_{ij} - n u_{ij}$. 
By 
\cite[Lemma 4]{lafontaine:conformal}
\begin{equation}\label{eq:kerA}
\ker A = \ker \tr \cap \ker d^{\nabla} .
\end{equation}
An element of $\ker d^{\nabla}$ is called a {\bf Codazzi tensor}, 
i.e. the kernel of $A$ consists of the trace--free Codazzi tensors. Clearly, a
trace-free Codazzi tensor is also a $\TT$--tensor. 
Let $\DD $ denote the Frechet derivative in the direction 
$$
(h,p) \in T_{(\gamma, -\gamma)} \CSlice_{\tau} ,
$$
as in section \ref{sec:geom-slice}.
\begin{lemma}\label{lem:DEB}
\begin{subequations}\label{eq:DEB}
\begin{align}
2\DD E (\gamma,-\gamma)(h,p) &=  Ah  -  ( n - 2 ) h - 2(n-2) p , 
\label{eq:DCperpperp}\\
\DD F (\gamma,-\gamma)(h,p) &= d^{\nabla}(p + h) . \label{eq:DCperp}
\end{align}
\end{subequations}
\end{lemma}
\begin{proof} 
By Lemma \ref{lem:TSlice}, $(h,p)$ are
$\TT$--tensors w.r.t. $\gamma$.
If $h$ is a $\TT$--tensor, w.r.t. $\gamma$, then 
\begin{equation}\label{eq:DRij}
\DD R_{ij}.h = \half \nabla^* \nabla h_{ij} - n h_{ij} = - \half \nabla^k
\nabla_k h_{ij} - n h_{ij} , 
\end{equation}
which gives (\ref{eq:DCperpperp}) after simplification. 
The Frechet derivative of $\Gamma^i_{jk}$ is given by 
\begin{equation}\label{eq:DGamma}
D \Gamma^i_{jk}. h 
= \half g^{im} ( \nabla_j h_{km} + \nabla_k h_{jm} - \nabla_m
h_{jk} )  .
\end{equation}
A computation using (\ref{eq:DGamma}) and (\ref{eq:CmTij}) yields
$$
\DD C_{mT ij}  
= \nabla_j h_{im} - \nabla_i h_{jm}  + \nabla_j p_{im} - \nabla_i p_{jm} ,
$$
which gives
(\ref{eq:DCperp}). 
\end{proof}
Consider a curve $\ame_\lambda$ of vacuum metrics on $\aM$, 
$\ame_0 = \agamma$, such that $\{M_t\}$ is CMCSH foliation with respect to
$\ame_{\lambda}$, and let $g_{\lambda}, k_{\lambda}$ be the induced data on
$M_{-n}$. Then 
$$
(h,p) = \frac{\partial}{\partial \lambda} 
(g_{\lambda} , k_{\lambda}) \bigg{|}_{\lambda = 0} ,  
$$
satisfy $(h,p) \in T_{(\gamma, - \gamma)} \CSlice_{-n}$. As above, let $\DD $
denote the Frechet derivative in the direction $(h,p)$. 
If we further assume that $\ame_{\lambda}$ is a family of flat metrics, 
then 
$\DD  E = 0$ and $\DD  F =0$. 

Decompose $h,p$ using the $L^2$--orthogonal direct sum decomposition 
$\ker A \oplus \ker^{T} A $, and write $h = h^0 + h^1$, $p= p^0 + p^1$, 
with $h^0, p^0 \in \ker A$, $h^1, p^1 \in \ker^{T}A$.
Then $\DD E = 0$ is equivalent to the system of equations
\begin{align}
  (n-2)[ h^0 +  2 p^0 ] &= 0 , \label{eq:kereq}\\
 Ah^1  -  ( n - 2   ) h^1
- 2(n-2) p^1 &= 0 . \label{eq:Aeq} 
\end{align}
By Lemma \ref{lem:TSlice}, $h,p$ are $\TT$--tensors on $(M,\gamma)$. 
Therefore,  by (\ref{eq:DCperp}), 
$h+p \in \ker \tr \cap \ker d^{\nabla}$, and hence, by 
(\ref{eq:kerA}), $h^1 + p^1 = 0$. 
This  means that equation (\ref{eq:Aeq}) is equivalent to 
$$
0 = A h^1 +  (n-2) h^1 .
$$
The restriction of $A$ to $\ker^{T}A$ is positive definite, so it 
follows that $h^1 = 0$. However, we know that $h^1 + p^1 = 0$, and hence 
$h^1 = p^1 = 0$.  Thus we have shown that $(h,p) = (h^0, p^0)$. 
It remains to make use of 
(\ref{eq:kereq}). In case $n=2$, this is trivial, while if $n \geq 3$, $h^0 +
2 p^0 = 0$ follows. 

By construction, $\ker \DD E \cap \ker \DD F$ is precisely the formal
tangent space $T_{\agamma} \Flat(\aM)$ at $\agamma$, of the space of 
flat Lorentz metrics $\Flat(\aM)$ on
$\aM$. 
Recalling that in dimension $2$, $\TT$--tensors are precisely trace--free
Codazzi tensors \cite{andersson:etal:2+1grav}, we have proved 
\begin{lemma} If $n=2$, $T_{\agamma} \Flat(\aM)$
is isomorphic to the direct sum of the space of $\TT$-tensors on $M$ with
itself,  while for $n\geq 3$, $T_{\agamma} \Flat (\aM)$ is isomorphic to the
space of trace--free Codazzi tensors on $M$. 
\qed
\end{lemma}  
In case $n=2$, $M$ is a Riemann surface of genus $\geq 2$, and in this case 
$T_{\agamma} \Flat(\aM)$ has dimension $12 \text{genus}(M) - 12$, while for
$n \geq 3$, $T_{\agamma} \Flat (\aM)$ is trivial in case $(M, \gamma)$  has no
non--vanishing trace--free Codazzi tensors, a topological condition. 
This motivates the following
definition. 
\begin{definition}\label{def:rigid-alt} A hyperbolic manifold $(M,\gamma)$
of dimension $3$, is 
{\bf rigid} if it admits no non--zero Codazzi tensors with vanishing trace. 
A hyperbolic cone space--time $(\aM,\agamma)$ is called rigid if $(M,\gamma)$ is
rigid. 
\end{definition}

A computation \cite{lafontaine:conformal}
shows that $(M,\gamma)$ is rigid in the sense of Definition
\ref{def:rigid-alt}, if and only if the formal tangent space at $\gamma$, 
of the space of
flat conformal structures on $M$ is trivial.  
Kapovich \cite[Theorem 2]{kapovich:deform} proved the existence
of compact hyperbolic $3$--manifolds which are rigid w.r.t. infinitesimal 
deformations in
the space of flat conformal structures.
We formulate this as 
\begin{prop}\label{prop:rigidexist-alt}
The class of rigid hyperbolic $3$--manifolds $(M,\gamma)$ (and rigid standard
space--times $(\aM, \agamma)$), in the sense of 
Definition \ref{def:rigid-alt}, is non--empty.
\qed
\end{prop}

\section{Weyl fields\label{sec:weyl}}
In this section, and in the rest of the paper, let $n = 3$. 
A tracefree 4-tensor $W$ with the symmetries of the Riemann tensor 
is called a Weyl field. We define the left and right Hodge duals of $W$ by
\begin{align}
\starW_{\alpha\beta\gamma\delta} &= \half \Eps_{\alpha\beta\mu\nu} W^{\mu\nu}_{\ \ \gamma\delta} ,\\ 
\Wstar_{\alpha\beta\gamma\delta} &= W_{\alpha\beta}^{\ \ \mu\nu} \half \Eps_{\mu\nu\gamma\delta} .
\end{align}
If $W$ is a Weyl field, then $\starW = \Wstar$ and $W = - *(\starW)$. 
Define the tensors $J$ and $J^*$ by 
\begin{subequations} \label{eq:inhomog-Bianchi}
\begin{align}
\anabla^\alpha W_{\alpha\beta\gamma\delta} &= J_{\beta\gamma\delta} , \\
\anabla^a {^*W}_{\alpha\beta\gamma\delta} &= J^*_{\beta\gamma\delta} .
\end{align}
\end{subequations}
Then 
$$
J^*_{\beta\gamma\delta} = \half J_\beta^{\ \mu\nu} \Eps_{\mu\nu\gamma\delta} ,
$$
and
\begin{subequations}\label{eq:fullBianchieqs}
\begin{align}
\anabla_{[\mu}W_{\gamma\delta]\alpha\beta} &=
\third \Eps_{\nu\mu\gamma\delta}J^{*\nu}_{\ \ \alpha\beta} ,
\label{eq:fullBianchi} \\
\anabla_{[e}\starW_{\gamma\delta]\alpha\beta} &=
-\third \Eps_{\nu\mu\gamma\delta}J^{\nu}_{\ \ \alpha\beta} .
\label{eq:fullBianchi*} 
\end{align} 
\end{subequations}
The electric and magnetic parts $E(W)$, $B(W)$ of the Weyl field $W$, with
respect to the foliation $M_t$ are
defined by 
\begin{equation}\label{eq:EBdef}
E(W)_{\alpha\beta} = W_{\alpha\mu\beta\nu}T^\mu T^\nu, \qquad B(W)_{\alpha\beta} = \starW_{\alpha\mu\beta\nu}T^\mu T^\nu .
\end{equation}
The tensors $E$ and $B$ are $t$--tangent, i.e. 
$E_{\alpha\beta}T^\beta = B_{\alpha\beta} T^\beta = 0$ and tracefree, 
$\ame^{\alpha\beta} E_{\alpha\beta} =
\ame^{\alpha\beta}B_{\alpha\beta} = 0$. It follows that $g^{ij} E_{ij} = g^{ij} B_{ij} = 0$. 

In case $(\aM, \ame)$ is vacuum, i.e. $\aR_{\alpha\beta}=0$, the Weyl tensor
$C_{\alpha\beta\gamma\delta}$ of $(\aM, \ame)$ satisfies $C_{\alpha\beta\gamma\delta} = \aR_{\alpha\beta\gamma\delta}$ the Gauss and
Codazzi equations can be written in terms of $E$ and $B$ to give
\begin{subequations}\label{eq:EBvac}
\begin{align}
\nabla_i k_{jm} - \nabla_j k_{im}   &= \Eps_{ij}^{\ \ l} B(\Weyltens)_{lm} ,
\label{eq:EpsB} \\
R_{ij} - k_{im} k^m_{\ j} + k_{ij} \tr k &= E(\Weyltens)_{ij} .
\end{align}
\end{subequations}
Note that from the definition (\ref{eq:extderdef}) 
of $d^{\nabla}$, (\ref{eq:EpsB}) is equivalent to 
$d^{\nabla} k_{mij} = - \Eps_{ij}^{\ \ l} B_{lm}$.
Using the definition (\ref{eq:symcurl}) of $\curl $ and 
(\ref{eq:dnabla-curl-div})
we get the alternate 
form of (\ref{eq:EpsB}),  valid if $(g,k)$ satisfies the vacuum 
constraint equations (\ref{eq:constraint}), 
\begin{equation}\label{eq:curlKB}
- (\curl k)_{ij} = B(\Weyltens)_{ij} .
\end{equation}
The following identities relate $W, \starW, E = E(W), B = B(W)$, 
cf. \cite[eq. (7.2.1), p. 169]{christo:klain}
\begin{equation}\label{eq:WEB}
\begin{aligned}
W_{ijkT} &= - \Eps_{ij}^{\ \ m} B_{mk} ,  &\quad  \starW_{ijkT} &= 
\Eps_{ij}^{\ \ m} E_{mk} , \\
W_{ijk\ell} &= - \Eps_{ijm} \Eps_{k\ell n} E^{mn} , &\quad \starW_{ijk\ell} &= 
- \Eps_{ijm}\Eps_{k\ell n} B^{mn} .
\end{aligned}
\end{equation}
The tensors 
$\anabla_T E$, $\anabla_T B$ have the property 
$$
g^{ij} \anabla_T E_{ij} = 0, \qquad g^{ij} \anabla_T B_{ij} = 0 ,
$$
i.e. the pull--back of $\anabla_T E$, $\anabla_T B$ to $M$ is trace--free,
but $\anabla_T E$, $\anabla_T B$ are not $t$--tangent in general.
The following result allows us to express derivatives of the Weyl field $W$
in terms of  $E(W), B(W), J(W), J^*(W)$. See Appendix \ref{sec:operations}
for the definition of $\div$ and $\curl$. 
\begin{prop}\label{prop:divcurlEB}
Let $E, B$ be the electric and magnetic parts 
of a Weyl-field $W$ and let $J, J^*$ be defined from $W$ by 
(\ref{eq:fullBianchieqs}). Then 
\begin{subequations}\label{eq:divEB}
\begin{align}
\div E_i &= + (k \wedge B)_i + J_{T iT} , 
\label{eq:divE}\\
\div B_i &= - (k \wedge E)_i + J^*_{T i T} . 
\label{eq:divB} 
\end{align}
\end{subequations}
\begin{subequations}\label{eq:curlEB}
\begin{align}
\anabla_T E_{ij} - \curl B_{ij} &= 
- \Lapse^{-1} ( \nabla \Lapse \wedge B)_{ij} 
- \frac{3}{2} ( E \times k)_{ij} + \half (\tr k) E_{ij} - J_{iT j } ,
\label{eq:curlB} \\
\anabla_T B_{ij} + \curl E_{ij}  &= 
+ \Lapse^{-1} ( \nabla \Lapse \wedge E)_{ij}  
- \frac{3}{2}(B \times k)_{ij} + \half (\tr k) B_{ij} - J^*_{iT j } .
\label{eq:curlE} 
\end{align}
\end{subequations}
Written in terms of $\Lie_{\partial_t}$, (\ref{eq:curlEB}) becomes
\begin{subequations}\label{eq:dtEB} 
\begin{align}
\Lapse^{-1} \Lie_{\partial_t} E_{ij} &= + \curl B_{ij} 
- \Lapse^{-1}(\nabla \Lapse \wedge
B)_{ij} \nonumber \\
&\quad - \frac{5}{2} (E \times k)_{ij} - \frac{2}{3} (E \cdot k) g_{ij} 
- \half \tr k E_{ij} + \Lapse^{-1} \Lie_{\Shift} E_{ij} -  J_{iT j} , \\
\Lapse^{-1} \Lie_{\partial_t} B_{ij} &= - \curl E_{ij} + \Lapse^{-1} (\nabla
\Lapse \wedge E)_{ij}  \nonumber \\
&\quad 
- \frac{5}{2} (B \times k)_{ij} - \frac{2}{3} (B \cdot k) g_{ij} - \half (\tr
k) B_{ij} + \Lapse^{-1} \Lie_{\Shift} B_{ij}  - J^*_{iT j } . 
\end{align}
\end{subequations}
\end{prop}
\begin{proof}We write $\anabla^{\alpha} E_{\alpha i}$ in two ways. First, 
$$
\div E_i = \anabla^{\alpha} E_{\alpha i}
-  \Lapse^{-1} \nabla^j \Lapse E_{ji} .
$$
Secondly, by (\ref{eq:WEB}) and (\ref{eq:symsymwedge}), 
\begin{align*}
\anabla^{\alpha} E_{\alpha i} &= 
\anabla^{\alpha} W_{\alpha T i T } 
+ \ame^{\alpha\beta} W_{\alpha \gamma i \delta } \anabla_{\beta} T^{\gamma}
T^{\delta} 
 + \ame^{\alpha\beta} W_{\alpha \gamma i \delta }
T^{\gamma}  \anabla_{\beta} T^{\delta}
\\
&= J_{T i T} + ( k \wedge B)_i + \Lapse^{-1} \nabla^j \Lapse E_{ji} .
\end{align*}
This gives (\ref{eq:divE}) and the argument for (\ref{eq:divB}) is similar. 
To prove (\ref{eq:curlEB}), first note the 
identities
\begin{subequations}\label{eq:anablacW}
\begin{align}
\anabla_k W_{iT jT } &= \nabla_k E_{ij} - 
( \Eps_{il}^{\ \ m} B_{mj} + \Eps_{jl}^{\ \ m} B_{mi}) k^l_{\ k} ,
\label{eq:anablacWab}\\
\anabla_k \starW_{iT jT } &= \nabla_k B_{ij} + 
( \Eps_{il}^{\ \ m} E_{mj} + \Eps_{jl}^{\ \ m} E_{mi}) k^l_{\ k} .
\label{eq:anablac*Wab}
\end{align}
\end{subequations}
From this we get, after expanding the covariant derivative, and rewriting using
(\ref{eq:WEB})  
\begin{subequations}\label{eq:symWeqs}
\begin{align} 
\Eps_i^{\ mn} \anabla_n W_{mT jT } + \Eps_j^{\ mn} \anabla_n
W_{mT iT } 
&= 2(\curl E)_{ij} + 3 ( B \times k)_{ij} - (\tr k) B_{ij}  , \label{eq:symW}\\
\Eps_i^{\ mn} \anabla_n \starW_{mT jT } + \Eps_j^{\ mn} \anabla_n
\starW_{mT iT } &= 2(\curl B)_{ij} 
- 3 ( E \times k)_{ij} + (\tr k) E_{ij} .
\label{eq:symW*}
\end{align}
\end{subequations}
The Bianchi equations (\ref{eq:fullBianchieqs}) imply
\begin{subequations}\label{eq:useBianchieqs}
\begin{align}
\Eps_i^{\ mn}\anabla_T  W_{mnjT } 
&= 2\Eps_i^{\ mn}\anabla_n W_{mT jT } - 2 J^*_{ijT } , \label{eq:useBianchi}\\
\Eps_i^{\ mn}\anabla_T  \starW_{mnjT } &= 
2\Eps_i^{\ mn}\anabla_n \starW_{mT jT } + 2 J_{ijT } .
\label{eq:useBianchi*}
\end{align}
\end{subequations}
Using (\ref{eq:WEB}) and (\ref{eq:framerel}) we get 
\begin{subequations} \label{eq:nabla0EB}
\begin{align}
\Eps_i^{\ mn}\anabla_T  W_{mnjT } 
&= - 2 \anabla_T  B_{ij} + 2 \Lapse^{-1} (\nabla \Lapse \wedge E)_{ij} ,
\label{eq:nabla0B} \\    
\Eps_i^{\ mn}\anabla_T  \starW_{mnjT } &= 
2 \anabla_T  E_{ij} 
+ 2 \Lapse^{-1} (\nabla \Lapse \wedge B)_{ij} . \label{eq:nabla0E}
\end{align}
\end{subequations}
Using (\ref{eq:nabla0EB}), multiplying by $\half$, taking the symmetric 
parts of (\ref{eq:useBianchieqs}) and using (\ref{eq:symWeqs}) now gives the
identities (\ref{eq:curlEB}). It is straightforward to derive 
(\ref{eq:dtEB}) from (\ref{eq:curlEB}) using (\ref{eq:Liedt}).
\end{proof}

Given a Weyl field $W$, the covariant derivative $\anabla_T W$ is again a
Weyl field. 
Proposition \ref{prop:divcurlEB} gives the following expressions for 
$E(\anabla_T W), B(\anabla_T W)$.
\begin{cor}\label{cor:EW1BW1}
\begin{subequations}\label{eq:EW1BW1}
\begin{align}
E(\anabla_T W)_{ij} &=  + \curl B_{ij} 
- \frac{3}{2} (E \times k)_{ij} + \half (\tr k ) E_{ij}  
- J_{iT j} ,
\label{eq:EW1}\\
B(\anabla_T W)_{ij} &= - \curl E_{ij} 
- \frac{3}{2} (B \times k )_{ij} + \half (\tr k) B_{ij} 
- J^*_{iT j} .
\label{eq:BW1} 
\end{align}
\end{subequations}
where in the right hand side, $E,B,J,J^*$ are defined w.r.t. $W$.
\end{cor}
\begin{proof}
From the definition and using (\ref{eq:framerel}) we have, taking into
account the fact that $E$ is $t$--tangent,
\begin{align*}
E(\anabla_T W)_{ij} &= T^\gamma T^\delta T^\nu \anabla_\nu W_{i\gamma j\delta} \nonumber \\
&= T^\nu \anabla_\nu E(W)_{ij} - \Lapse^{-1}\nabla^m \Lapse  W_{imjT } 
-  \Lapse^{-1}\nabla^n \Lapse  W_{iT jn} \nonumber \\ 
\intertext{using (\ref{eq:WEB}) and (\ref{eq:vecsymwedge})}
&= T^\nu \anabla_\nu E(W)_{ij} + \Lapse^{-1} (\nabla \Lapse \wedge B(W))_{ij} ,
\end{align*}
which gives (\ref{eq:EW1}) using (\ref{eq:curlB}). The proof of (\ref{eq:BW1})
is similar. 
\end{proof}

\section{The Bel-Robinson Energy\label{sec:BRenergy}}
Given a Weyl field $W$ we can associate to it a fully symmetric and
traceless tensor 
\begin{equation}
\begin{split}
\BR(W)_{\alpha\beta\gamma\delta} &=  
W_{\alpha\mu\gamma\nu}W_{\beta\ \delta}^{\ \mu\ \nu} + 
\starW_{\alpha\mu\gamma\nu}\starW_{\beta\ \delta}^{\ \mu\ \nu} . 
\end{split}
\end{equation}
$\BR(W)$ is positive definite in the sense that $\BR(X,Y,X,Y) \geq 0$ 
whenever $X,Y$ are timelike vectors, with equality only if $W$ vanishes, cf. 
\cite[Prop. 4.2]{christo:klain:linear} 
Let $E = E(W), B = B(W)$. The following identities relate $\BR(W)$ to $E$ and 
$B$.
\begin{subequations}\label{eq:BRid}
\begin{align}
\BR(W)_{T T T T } &= |E|^2 + |B|^2 , 
\label{eq:BREB}\\ 
\BR(W)_{iT T T } &= 
2 (E \wedge B)_i , \label{eq:BREBi000} \\
\BR(W)_{ijT T } &= 
-  (E \times E)_{ij} -  ( B \times B)_{ij} + \third (|E|^2 + |B|^2) g_{ij} ,
\label{eq:BREBij00}
\end{align}
\end{subequations}
where $|E|^2 = E^{ij} E_{ij} = |E|^2_g$, and similarly for $|B|^2$. 
From equations (\ref{eq:WEB}) and (\ref{eq:BREB}) it follows that 
$\BR(W)_{T T T T }=0$ if and only if $W = 0$. 
The divergence of the Bel--Robinson tensor takes the form 
\cite[Prop. 7.1.1]{christo:klain}
\begin{equation}\label{eq:DivQ}
\begin{split}
\anabla^{\alpha} \BR(W)_{\alpha\beta\gamma\delta} &=
W_{\beta\ \delta}^{\ \mu\ \nu}J(W)_{\mu\gamma\nu} 
+ W_{\beta\ \gamma}^{\ \mu\ \nu}J(W)_{\mu\delta\nu} \\
&\quad 
+ \starW_{\beta\ \delta}^{\ \mu\ \nu}J^*(W)_{\mu\gamma\nu} 
+ \starW_{\beta\ \gamma}^{\ \mu\ \nu}J^*(W)_{\mu\delta \nu}  ,
\end{split}
\end{equation}
and the definition of $E(W)$ and $B(W)$ gives
\begin{equation}\label{eq:DivQ000}
\anabla^\alpha \BR(W)_{\alpha T T T } 
=2 E^{ij}(W) J(W)_{iT j} + 2 B^{ij}(W) J^*(W)_{iT j} .
\end{equation}
Let $W$ be a Weyl field and let $\BR(W)$ be the corresponding Bel-Robinson
tensor. Then working in a foliation $M_t$, 
we define the Bel-Robinson energy $\EBR(t,W)$ by 
$$
\EBR(t,W) = \int_{M_t} \BR(W)_{T T T T } d\mu_{M_t} .
$$
By the Gauss law, 
this has the evolution equation 
\begin{equation}\label{eq:BRcons}
\begin{aligned}
\partial_{t } \EBR(t,W) &= 
- \int_{M_t} \Lapse \anabla^\alpha \BR(W)_{\alpha T T T } d\mu_{M_t}  \\
& \quad 
- 3 \int_{M_t}\Lapse \BR(W)_{\alpha\beta T T }
\pi^{\alpha\beta}  d\mu_{M_t} ,
\end{aligned}
\end{equation}
where $\pi$ denotes the 
``deformation tensor'' of $T$, 
\begin{equation}\label{eq:pi-def}
\pi_{\alpha\beta} = \anabla_{\alpha}T_{\beta} .
\end{equation}
The components of $\pi$ in terms of an adapted, Fermi--propagated frame are 
as follows: 
\begin{subequations}\label{eq:picomp}
\begin{align}
\pi_{ij} &=  - k_{ij} , &
\pi_{iT} &= 0 , \\
\pi_{T i} &= \Lapse^{-1} \nabla_i \Lapse , &
\pi_{T T } &= 0 .
\end{align}
\end{subequations}
We will need 
control of $\ame$ in $H^3$, and for this purpose we consider in addition to
the Bel--Robinson energy of order zero, $\EBR_0(t,W)= \EBR(t,W)$, 
the first order Bel--Robinson energy 
$\EBR_1(t,W) = \EBR(t,\anabla_T W)$. 
In the vacuum case, $J(W) = J^*(W) = 0$, so 
we may view $\EBR_1$ as a function on the
set of solutions to the Einstein vacuum constraint equations, by using
Corollary \ref{cor:EW1BW1} to compute $E(\anabla_T W), B(\anabla_T W)$.  

Expanding $\partial_t \EBR(t,W)$ using (\ref{eq:BRid}), (\ref{eq:DivQ000}), 
(\ref{eq:picomp}),
gives 
\mnote{changed sign in front of $\Lapse^{-1} \nabla^i \Lapse  (E \wedge
B)_i$, check it}
\begin{align}
\partial_t Q(t,W) &= 
- 3 \int_{M_t} \Lapse [(E\times E)\cdot k  + (B\times B)\cdot k  
 - \third (|E|^2 + |B|^2)\tr k \nonumber \\
&\quad - 2 \Lapse^{-1} \nabla^i \Lapse  (E \wedge B)_i ] d\mu_{M_t} \nonumber
\\
&\quad - 2 \int_{M_t} \Lapse ( E^{ij} J_{iT j} + B^{ij} J^*_{iT j}) d\mu_{M_t}
\nonumber \\
\intertext{(perform a partial integration and use (\ref{eq:divAwedgeB}))}
&= - 3 \int_{M_t} \Lapse [ (E \times E) \cdot k + (B \times B) \cdot k 
-\third (|E|^2 + |B|^2) \tr k \nonumber \\
&\quad - 2 \curl E \cdot B + 2 E \cdot \curl B ] d\mu_{M_t} \nonumber \\
&\quad - 2 \int_{M_t} \Lapse ( E^{ij} J_{iT j} + B^{ij} J^*_{iT j })
d\mu_{M_t} . \label{eq:dtQ}
\end{align}
It is 
\mnote{put in an equation number, but we are not referring to it in the paper} 
straightforward to show that this expression agrees
with that obtained after a direct computation of $\partial_t Q(t,W)$ 
using (\ref{eq:dtEB}). 

Let $\tau = \tr k$ and specialize to a constant mean curvature foliation
$\{M_\tau\}$ in the following. 
With the discussion in subsection \ref{sec:lobell} as a guide we introduce the following 
quantities which vanish when evaluated in the standard foliation on
a hyperbolic cone space--time, namely
the ``trace free'' part 
$\hpi$ of $\pi$,
\begin{align}
\hpi_{\alpha\beta} &= \pi_{\alpha\beta} 
+ \frac{\tau}{3} (\ame_{\alpha\beta} + T_{\alpha} T_{\beta} ) 
\label{eq:hpi-def} \\
\intertext{and the ``perturbed'' part of the lapse, }
\hLapse &= \Lapse - \frac{3}{\tau^2} . \label{eq:hLapsedef}
\end{align}


In the following Lemma we record the form of $\partial_\tau \EBR_i$, $i=0,1$
which will be used in the global existence proof. 

\begin{lemma}\label{lem:cons-laws}
In a vacuum space time, the Bel--Robinson energies $\EBR_j(t,W)$, $j=0,1$ satisfy
the following conservation laws.
\begin{equation}\label{eq:BRconsJ=0}
\begin{split}
\partial_\tau \EBR_0(\tau,W) &= \frac{3}{\tau} \EBR_0(t,W) 
- 3 \int_{M_\tau}\Lapse \BR(W)_{\alpha\beta\gamma\delta}
\hpi^{\alpha\beta} T^{\gamma} T^{\delta} d\mu_{M_\tau} \\
&\quad 
+ \tau \int_{M_\tau}
\hLapse \BR(W)_{TTTT}
d\mu_{M_\tau} ,
\end{split}
\end{equation}

\begin{equation}\label{eq:BRconsJi}
\begin{split}
\partial_\tau \EBR_1(\tau,W) &= \frac{5}{\tau} \EBR_1(t,W) 
- 2 \int_{M_{\tau}} N \GG_1(W)
d\mu_{M_\tau} \\
& \quad 
- 3 \int_{M_t}\Lapse \BR(\anabla_T W)_{\alpha\beta\gamma\delta}
\hpi^{\alpha\beta} T^{\gamma} T^{\delta} d\mu_{M_\tau} \\
&\quad 
+ \frac{5\tau}{3} \int_{M_\tau}
\hLapse \BR(\anabla_T W)_{TTTT}
d\mu_{M_\tau} ,
\end{split}
\end{equation}
where 
\begin{equation}\label{eq:GG-general}
\begin{split}
\GG_1(W) &= E(\anabla_T W )^{ij} ( J(\anabla_T W)_{iT j} + \frac{\tau}{3} E(\anabla_T W)_{ij} )  \\
&\quad + B(\anabla_T W)^{ij}(J^*(\anabla_T W)_{iT j} + \frac{\tau}{3} 
B(\anabla_T W)_{ij}) . 
\end{split}
\end{equation}
In particular,
\begin{subequations} \label{eq:J-details}
\begin{align}
J(\anabla_T W)_{iT j} + \frac{\tau}{3}  E(\anabla_T W)_{ij}    &= 
\hpi^{\alpha\mu}\anabla_\mu W_{\alpha iT j} 
+\frac{3}{2} (E \times E)_{ij} - \frac{3}{2} (B \times B)_{ij}
\label{eq:J-details-W} \\
J^*(\anabla_T W)_{iT j} + \frac{\tau}{3} B(\anabla_T W)_{ij}   &= 
\hpi^{\alpha\mu}\anabla_\mu \starW_{\alpha iT j} 
+ 3 (E \times B)_{ij} 
\label{eq:J-details-*W} 
\end{align}
\end{subequations}
where 
\begin{subequations}\label{eq:pinablaW-details}
\begin{align}
\hpi^{\alpha\mu}\anabla_\mu W_{\alpha iT j} &= \hk^{rs} \nabla_s
(-\eps_{ri}^{\ \ n}B_{nj} ) \nonumber \\
&\quad - \hk^{rs} k_{rs} E_{ij} + \hk^{rs} k_{si} E_{rj} + \hk^{rs} k_s^{\ m}
\eps_{ri}^{\ \ n} \eps_{mj}^{\ \ p} E_{np}
\nonumber \\
&\quad - \Lapse^{-1} \nabla^r \Lapse ( \nabla_r E_{ij} 
+ k_r^{\ s}\eps_{si}^{\ \ n}  B_{nj} + k_r^{\ s} \eps_{sj}^{\ \ n} B_{ni} ) \\
\hpi^{\alpha\mu}\anabla_\mu \starW_{\alpha iT j} &= \hk^{rs} \nabla_s
(\eps_{ri}^{\ \ n} E_{nj}) \nonumber \\
&\quad - \hk^{rs} k_{rs} B_{ij} + \hk^{rs} k_{si} B_{rj}
+ \hk^{rs} k_s^{\ m} \eps_{ri}^{\ \ n} \eps_{mi}^{\ \ p} B_{np}   
\nonumber \\
&\quad - \Lapse^{-1} \nabla^r \Lapse ( \nabla_r B_{ij} - k_r^{\ s} \eps_{si}^{\
\ n} E_{nj} - k_r^{\ s} \eps_{sj}^{\ \ n} E_{ni} ) 
\end{align}
\end{subequations}
\end{lemma}
\begin{remark}\label{rem:struct}
In the proof of the main theorem, it is of central importance that the terms
in (\ref{eq:J-details}) are quadratic in $\hpi, \anabla W, W$. This has the
consequence that the term given by (\ref{eq:GG-general}) can be treated as a
perturbation term in case of small data. In particular the terms 
$\hpi^{\alpha\mu}\anabla_\mu W_{\alpha iT j} $ and 
$ \hpi^{\alpha\mu}\anabla_\mu \starW_{\alpha iT j}$ when expanded are
seen to be of third order in $N^{-1} \nabla_i N, k_{ij}, E_{ij}, B_{ij}$ 
and of  second order in $\hk_{ij}, \nabla_i E_{jk}, \nabla_i B_{jk}$. 
We will not make use of the explicit expression for $\partial_t \EBR_1$, but
for completeness, it is given in equation (\ref{eq:Vince-big-dtEBR}) below.
\end{remark}
\begin{proof}
In order to evaluate $\Div \BR(\anabla_T W)(T,T,T)$, we need $J(\anabla_T W)_{iT j}$ and
$J^*(\anabla_T W)_{iT j}$.
A computation gives 
\begin{equation}\label{eq:weylone-div}
\begin{aligned}
J(\anabla_T W)_{\beta\gamma\delta} = \anabla^\alpha \anabla_T W_{\alpha\beta\gamma\delta} &= \pi^{\alpha\nu} \anabla_\nu W_{\alpha\beta\gamma\delta} + 
T^\nu \anabla_\nu J(W)_{\beta\gamma\delta}  \\
&\quad + T^\nu \aR_{\alpha\ \ \nu}^{\ \mu\alpha} W_{\mu\beta\gamma\delta} 
+ T^\nu \aR_{\beta\ \ \nu}^{\ \mu\alpha} W_{\alpha\mu\gamma\delta} \\
&\quad 
+ T^\nu \aR_{\gamma\ \ \nu}^{\ \mu\alpha} W_{\alpha\beta\mu\delta}
+ T^\nu \aR_{\delta\ \ \nu}^{\ \mu\alpha} W_{\alpha\beta\gamma\mu} ,
\end{aligned}
\end{equation}
Note that in vacuum, $J(W) = 0$ and 
$\aR_{\alpha\beta\gamma\delta} = W_{\alpha\beta\gamma\delta}$. 
Substituting $\aR$ for $W$ in (\ref{eq:weylone-div}) 
and using (\ref{eq:WEB}) to
rewrite the terms quadratic in $W$ gives (\ref{eq:J-details-W}). 
A similar calculation for $J^*(\anabla_T W)$, taking into account the fact
that in this case, $\aR=W$ and $\starW$ are distinct,
gives (\ref{eq:J-details-*W}). 
It is now straightforward to check that 
(\ref{eq:BRconsJ=0}) and (\ref{eq:BRconsJi}) hold, given the definition of
$\GG_1$ in (\ref{eq:GG-general}). 
\end{proof}

One can decompose equations (\ref{eq:J-details}) 
into symmetric and antisymmetric parts to obtain the
analogues of equations (\ref{eq:divEB}) and (\ref{eq:curlEB}). Setting $J(W)
= J^*(W) = 0$ for the vacuum case, defining $\tE_{ij} = E(\anabla_T W)_{ij}$
and $\tB_{ij} = B(\anabla_T W)_{ij}$ and writing $E_{ij|k}$ and $B_{ij|k}$ for
$\nabla_k E_{ij}$ and $\nabla_k B_{ij}$ respectively we get, first for the
symmetric parts (the analogues of equations (\ref{eq:dtEB})),

\begin{subequations}\label{eq:VincedtEB}
\begin{align}
\Lapse^{-1} \Lie_{\partial_t} \tE_{ij} &= 
k_{ij} ( k^{lm} E_{lm} ) 
+ g_{ij}  [ E^{lm} E_{lm} + k^{ls} k_s^{\ m} E_{lm} 
+ k^{lm} \tE_{lm} - B^{lm} B_{lm} ] 
\nonumber \\ &\quad 
+ 2E_{ij} ( k^{lm} k_{lm}) 
- 3 E_{il} E^l_{\ j} - 3 k_i^{\ m} \tE_{mj}
\nonumber \\ &\quad 
-3 k_j^{\ m} \tE_{mi} 
- k_i^{\ m} k_j^{\ l} E_{lm} + 3 B_{il} B^l_{\ j}
\nonumber \\
&\quad + \half g_{jk} \eps^{lkm} ( \tB_{im|l} - k_{ms} B^s_{\ i|l} ) 
+ \half g_{ik} \eps^{lkm} ( \tB_{jm|l} - k_{ms} B^s_{\ j|l} ) \nonumber \\
&\quad 
+ \Lapse^{-1} \Lapse^{|l} E_{ij|l} - \frac{5}{2} k_j^{\ s} k_s^{\ m} E_{im} 
- \frac{5}{2} k_i^{\ s} k_s^{\ m} E_{jm} \nonumber \\
&\quad + (\tr k) [ 3\tE_{ij} + \frac{5}{2} k_i^{\ m} E_{jm} + \frac{5}{2} k_j^{\ m}
E_{im} - 2 (\tr k) E_{ij} - g_{ij} k^{lm} E_{lm} ] \nonumber \\
&\quad + \Lapse^{-1} \Lapse^{|l} [ B^r_{\ i} \eps_{kjr} k_l^{\ k} + B^r_{\ j}
\eps_{kir} k_l^{\ k} - \eps_{rjl} \tB^r_{\ i} - \eps_{ril} \tB^r_{\ j} ]
\nonumber \\
&\quad + \Lapse^{-1} \Lie_{\Shift} \tE_{ij} \label{eq:Vince-dtE} \\
 & \quad \nonumber \\
\Lapse^{-1} \Lie_{\partial_t} \tB_{ij} &= 2 B_{ij} k^{lm} k_{lm} + g_{ij} [
2B^{lm} E_{lm} + k^{ls} k_s^{\ m} B_{lm}  + k^{lm} \tB_{lm} ] \nonumber \\
&\quad + \Lapse^{-1} \Lapse^{|l}B_{ij|l} 
- 3 B^m_{\ j} E_{im} - 3 B^m_{\ i} E_{jm} - 3 k_i^{\ l} \tB_{lj} \nonumber \\
&\quad   
-3k_j^{\ l} \tB_{li} - k_i^{\ l} k_j^{\ s} B_{ls} + k_{ij} k^{lm} B_{lm}
\nonumber \\
&\quad 
- \frac{5}{2} ( k_{jm} k^{ml} B_{li} + k_{im} k^{ml} B_{lj} ) \nonumber \\
&\quad - \half g_{ik} \eps^{lkm} ( \tE_{mj|l} - k_m^{\ s} E_{sj|l} ) 
- \half g_{jk} \eps^{lkm} ( \tE_{mi|l} - k_m^{\ s} E_{si|l} ) \nonumber \\
&\quad + ( \tr k) [ 3 \tB_{ij} + \frac{5}{2} k_i^{\ l} B_{lj} 
+ \frac{5}{2} k_j^{\ l} B_{li} - 2 (\tr k) B_{ij} - g_{ij} k^{lm} B_{lm} ] 
\nonumber \\
&\quad - \Lapse^{-1} \Lapse^{|l} [ E^r_{\ i} \eps_{kjr} k^k_{\ l} + E^r_{\ j}
\eps_{kir} k^k_{\ l} - \eps_{rjl} \tE^r_{\ i} - \eps_{ril} \tE^r_{\ j} ] 
\nonumber \\
&\quad 
+ \Lapse^{-1} \Lie_\Shift \tB_{ij} \label{eq:Vince-dtB}
\end{align}
\end{subequations}
and then for the antisymmetric part (the analogues for equations
(\ref{eq:divEB})),
\begin{subequations}\label{eq:Vince-divtEB}
\begin{align}
\tE^{i\ \ |j}_{\ j} 
&= k_j^{\ m} \tB_{mr} \eps^{ijr} - k_j^{\ m} E^{i\ \ |j}_{\ m} 
+ (\tr k) \eps^{imn} k_m^{\ s} B_{sn} - k^i_{\ r} k_l^{\ s} B_{sj} \eps^{rlj} \label{eq:Vince-divtE}
\\
\tB^{i\ \ |j}_{\ j} &= - k_j^{\ m} \tE_{mr} \eps^{ijr} 
- k_j^{\ m} B^{i\ \ |j}_{\ m} - (\tr k) \eps^{imn} k_m^{\ s} E_{sn} + k^i_{\
r} k_l^{\ s} E_{sj} \eps^{rlj} \label{eq:Vince-divtB}
\end{align} 
\end{subequations}

The formula for $\partial_t \EBR_1$ that is analogous to that given above for
$\partial_t \EBR$ is given explicitely as follows, 
%
\newcommand{\Ktr}{\hk}
%
\begin{align}
\frac{\partial}{\partial t} \int_M &  \mu_g ( \tE_i^\ell \tE_\ell^i 
+ \tB_i^\ell \tB_\ell^i ) = \partial_t \EBR_1 \nonumber \\
&= 2 \int_M \bigg\{ N g_{jk} \eps^{\ell k m} \left [ \Ktr^{\ s}_m ( \tB^{ij}
E_{si|\ell} - \tE^{ij} B_{si|\ell} )\right ]\mu_g 
\nonumber \\
&\quad 
+ N \mu_g ( \tr k) \left [ \frac{5}{6} \tE^{ij} \tE_{ij} +
\frac{5}{6}\tB^{ij}\tB_{ij} \right ] 
\nonumber \\
&\quad 
+\mu_g N^{|\ell} \left ( \tE^{ij} E_{ij|\ell} +  \tB^{ij} B_{ij|\ell} \right ) 
\nonumber  \\ 
&\quad 
+ 2N \mu_g \left ( \Ktr_{mn}\Ktr^{mn} \right ) 
\left ( \tE^{ij} E_{ij} + \tB^{ij} B_{ij} \right ) 
\nonumber  \\
&\quad -4 N\mu_g \Ktr_j^\ell \left [ \tE_\ell^i \tE_i^j + \tB_\ell^i \tB_i^j
\right ] 
\nonumber   \\
&\quad + N \mu_g \left[ ( \tE^{ij} \Ktr_{ij}  ) ( E^{mn} \Ktr_{mn}  ) 
+  ( \tB^{ij} \Ktr_{ij} ) ( B^{mn} \Ktr_{mn} ) \right ] 
\nonumber  \\
&\quad - 3 N \mu_g \left [ \tE^{ij} E_{i\ell} E^\ell_j 
- \tE^{ij} B_{i\ell} B^\ell_j + 2 \tB^{ij} B^\ell_j E_{i\ell} \right ]
\nonumber  \\
&\quad 
- 3 N_{|\ell} \eps^{rj\ell} \tE^i_j \tB_{ri} \mu_g 
+ 2N^{|\ell} \left [ \tE^i_j B_{ri} \eps^{kjr} ( \Ktr_{\ell k} + \third
g_{lk} ( \tr k) ) \right. 
\nonumber  \\
&\quad 
\left. 
- \tB^i_j E_{ri} \eps^{kjr} ( \Ktr_{k\ell} + \third g_{kl} (\tr k)  ) 
\right ]\mu_g 
\nonumber  \\
&\quad - N \mu_g \Ktr_i^m \Ktr_j^\ell \left[ \tE^{ij} E_{lm} + \tB^{ij}
B_{\ell m} \right ] 
\nonumber \\
&\quad 
- 5 N \mu_g \Ktr_j^s \Ktr_s^m ( \tE^{ij} E_{im} + \tB^{ij} B_{im} ) \bigg\}
\label{eq:Vince-big-dtEBR}   
\end{align}

\subsection{The scale--free Bel--Robinson energy}\label{sec:BRvac}
In the rest of section \ref{sec:BRenergy}, let $n=3$, 
and assume $(\aM, \ame)$ is a vacuum space--time
with a CMC foliation $\{M_\tau\}$ with $\tau = \tr k < 0$.
Let $W_{\alpha\beta\gamma\delta} = \aR_{\alpha\beta\gamma\delta}$ 
be the Weyl tensor of $(\aM, \ame)$. Then $W$
satisfies the homogenous Bianchi identities, i.e. $J(W) = J^*(W) = 0$. 

Since we will be estimating geometric
quantities in terms of $\EBR_0, \EBR_1$ via Sobolev inequalities which depend 
on scale, we need scale--free versions of these energies. 
It follows from the definitions that the following variables are 
scale--free if $\lambda$ has dimensions $\length^{-1}$. Here indices refer to
a coordinate frame. 
\begin{subequations}\label{eq:scaling}
\begin{align}
\tg_{ab} &= \lambda^2 g_{ab} , &
\tT^\alpha &= \lambda^{-1} T^\alpha , \\
\tk_{ab} &= \lambda k_{ab} , &
\widetilde{\tr k}&= \lambda^{-1} \tr k , \\
\tLapse &= \lambda^2 \Lapse , &
\widetilde{\mu_g} &= \lambda^3 \mu_g .
\end{align}
\end{subequations}
\mnote{\Lars check scaling of $\Lapse$} 
Note
that $\tr k$ has dimensions of $\length^{-1}$ while we treat spatial 
coordinates as dimensionless quantities.
The Weyl tensor in the (3,1) form is conformally invariant, 
and hence the Bel--Robinson tensor is also conformally invariant, and in
particular scale--free. 
From this can be seen that 
the Bel--Robinson energies $\EBR_0, \EBR_1$ have dimensions $\length^{-1}$ 
and $\length^{-3}$ respectively, so that the expressions 
$$
\tEBR_i = \lambda^{-1-2i} \EBR_i , \qquad i = 0,1,
$$
are scale--free, i.e. $\tEBR_i$ is precisely given by $\EBR_i$ evaluated on
the scale--free variables $\tg, \tk$ etc. 
In the following we will use the scale factor $\lambda$, defined 
by 
\begin{equation}\label{eq:lambda-def}
\lambda = \frac{|\tr k|}{3} = - \frac{\tr k}{3}.
\end{equation}

The scale--free energy function which will be used in the proof of global
existence is the sum of $\tEBR_0$ and $\tEBR_1$, 
\begin{equation}\label{eq:energy-def}
\Energy =  \tEBR_0  + \tEBR_1.
\end{equation}
It is convenient to introduce 
the logarithmic time $\logtime = - \ln(-\tau)$. Then 
$\logtime \nearrow \infty$ as $\tau \nearrow 0$.
The logarithmic time $\logtime$ has the property that 
$\partial_{\logtime} = - \tau \partial_\tau$ is scale--free, so that for example
$\partial_{\logtime} \Energy$ is scale--free.

\subsection{The Hessian of the Bel--Robinson energy}\label{sec:BRhess}
Let $M$ be a compact 3--dimensional manifold of hyperbolic type and let
$\gamma$ be the standard hyperbolic metric on $M$. 
Let 
$(\frac{9}{\tau^2} \gamma, \frac{3}{\tau} \gamma)$ be data with mean
curvature $\tau$ for the 
hyperbolic cone space--time $(\aM, \agamma)$. 
For $s \geq 3$, 
let $\CSlice_{\tau}^s$ be the local
slice for the action of $\Diff^{s+1}$ on the constraint set $\Constr_{\tau}$, 
at the hyperbolic cone data with mean curvature $\tau$. 

The energies $\tEBR_0, \tEBR_1, \Energy$ may be thought of as functions on 
the constraint set $\Constr_{\tau}^s$ by using equations (\ref{eq:EBvac}) and
(\ref{eq:WEB}). The following Lemma is a straightforward consequence of 
the Sobolev imbedding theorems.   
\begin{lemma}\label{lem:energy-regular}
The scale--free energies  $\tEBR_0(\tau,W), \tEBR_1(\tau,W), 
\Energy(\tau,W)$, are
$C^{\infty}$ functions on $\Constr_{\tau}^3$ and
$\CSlice_{\tau}^3$.
\end{lemma}
Let $\Hess \tEBR_0(\gamma, - \gamma)$ denote the Hessian of the function
$\tEBR_0$, evaluated at $(\gamma, - \gamma)$. 
A computation shows that for $(h,p) \in T_{(\gamma, - \gamma)}
T\CSlice_{-n}$, 
$$
\Hess \tEBR_0(\gamma,-\gamma)  ( ( h , p )
, 
 (h , p )) 
= \half || Ah||_{L^2}^2 + (Ap,p)_{L^2} + \half || h+2p||_{L^2}^2  ,
$$
where $A$ is given by (\ref{eq:Adef}), and $|| \cdot ||_{L^2}, (\cdot ,
\cdot)_{L^2}$ denote the $L^2$ norm and inner product defined with respect to
$\gamma$. Recall that $\ker A = \{0\}$ if and only if $M$ is rigid. 
It is now straightforward to prove the following Lemma.
\begin{lemma}\label{lem:EBRhess}
Let $M$ be a compact hyperbolic $3$--manifold. The Hessian of the scale--free
Bel--Robinson energy $\tEBR_0$, defined by equation 
(\ref{eq:energy-def}), considered as a function on 
$\CSlice_{-3}$, 
evaluated at the standard data $(\gamma,-\gamma)$, satisfies the inequality
\begin{equation}\label{eq:HessEBRineq-0}
\Hess \tEBR_0(\gamma,-\gamma) ( ( h , p), ( h , p )  )
\geq C ( ||h||^2_{H^2} + || p||^2_{H^1} ) ,
\end{equation}
if and only if $M$ is rigid. 
The constant $C$ depends only on the topology of $M$. 
\qed
\end{lemma}
Consider a solution $\ah$ of the linearized Einstein equations on the
hyperbolic cone space--time $(\aM, \agamma)$.
The derivative of the Weyl tensor in the direction
of $h$, 
$$ 
W' = DW[\agamma].\ah ,
$$
is a Weyl field on $(\aM, \agamma)$ which satisfies the homogeneous
Bianchi equations,
\begin{equation}\label{eq:JWprim}
J(W') = J^*(W') = 0 ,
\end{equation}
Let $E(W'), B(W')$ be the electric and magnetic parts
of $W'$ at $M_{-3} \subset \aM$. Then  
$$
DE(\anabla_T W) {|}_{\agamma} \ah = E(\anabla_T W'), \qquad 
DB(\anabla_T W) {|}_{\agamma} \ah = B(\anabla_T W').
$$
Recall that the second fundamental form of $M_{-3}$ is $k = - \gamma$.
This implies using (\ref{eq:divEB}), (\ref{eq:JWprim}), 
\begin{equation}\label{eq:EBWp}
\div E(\anabla_T W') = \div B(\anabla_T W') = 0 ,
\end{equation}
which shows that $E(\anabla_T W')$ and $B(\anabla_T W')$ are $\TT$--tensors. 
It follows from Corollary \ref{cor:EW1BW1}, using $k = - \gamma$ and 
(\ref{eq:JWprim}), 
\begin{align*}
E(\anabla_T W') &= + \curl B(W') + \half \tr k E(W') ,\\
B(\anabla_T W') &= - \curl E(W') + \half \tr k B(W')  .
\end{align*}
The scale--free Bel--Robinson
energy $\Energy$ is a smooth function on
$\CSlice_{\tau}^{3}$. 
Using the above it is straightforward, using (\ref{eq:quadeq}) 
to prove that the Hessian of
$\Energy$ is positive definite on $H^3 \times H^2$ in case $M$ is rigid. 
We state this as  
\begin{thm}\label{thm:EBRhess-2}
The hessian $\Hess \Energy$ on $\CSlice_{\tau}$, 
evaluated at the standard data 
$(\frac{\tau^2}{9} \gamma,\frac{\tau}{3} \gamma)$, satisfies the inequality
\begin{equation}\label{eq:HessEBRineq-2}
\Hess \Energy(\tfrac{9}{\tau^2}\gamma,\tfrac{3}{\tau} \gamma) 
(  (h , p ) , 
(h , p ) )
\geq C ( ||h||^2_{H^3} + || p||^2_{H^2} ) ,
\end{equation}
if and only if $M$ is rigid. The 
constant $C$ depend only on the topology of $M$. 
\end{thm}
Results analogous to Theorem \ref{thm:EBRhess-2} can easily be proved for
even higher order Bel--Robinson type energies. This will not be needed in
this paper.

\section{Estimates\label{sec:estimates}}
In this section we will introduce a ``smallness condition'' on the vacuum
data $(g,k)$, under which we are able to control all relevant geometric
quantities in terms of the energy function $\Energy$ defined in
section \ref{sec:BRvac}.
Recall the definition of the slice $\CSlice_{\tau}$ in section
\ref{sec:geom-slice}. In particular, vacuum data $(g,k)\in \CSlice_{\tau}$
satisfy the CMCSH gauge conditions. 
\begin{definition}\label{def:smallness}
Let $(g,k)$ be a vacuum data set on $M$ with mean curvature $\tau$. Let
$\lambda$ be given by (\ref{eq:lambda-def}) and let $(\tg,
\tk)$ be the rescaled metric and second fundamental form
as defined in (\ref{eq:scaling}). Let $\Smallset(\alpha)$ be the set 
of $(g,k) \in \CSlice_{\tau}^{3}$ so that 
$$
|| \tilde g - \gamma ||_{H^3}^2 + || \tilde{k} + \gamma ||_{H^2}^2 < \alpha .
$$
We will say that $(g,k)$ satisfies the smallness condition if $(g,k) \in
\Smallset(\alpha)$. 
\qed
\end{definition}
The smoothness of the scale--free Bel--Robinson energy $\Energy$, Lemma
\ref{lem:energy-regular}, together with
the fact that the Hessian of $\Energy$, 
the scale--free Bel--Robinson energy restricted
to the slice $\CSlice_{\tau}$ 
is positive definite with respect to the Sobolev norm $H^3 \times H^2$, 
Theorem \ref{thm:EBRhess-2}, implies, by
Taylor's theorem, the following estimate. 
\begin{thm}\label{thm:energy-norm} Assume that $M$ is rigid. 
There is an $\alpha > 0$ so that for $(g,k) \in \Smallset(\alpha)$,
there is a constant $D(\alpha) < \infty$, depending only on $\alpha$
and the topology of $M$, such that
\begin{align}
D(\alpha)^{-1} \Energy(g,k) \leq || \tilde g - \gamma ||_{H^3}^2 + || \tilde{k} +
\gamma||_{H^2}^2 \leq D(\alpha)\Energy(g,k) . 
\end{align}
\end{thm}
In view of the analysis of the elliptic defining equations for $\Lapse,
\Shift$ in 
\cite{andersson:moncrief:local}, there is a neighborhood of 
$( \gamma , - \gamma)$ in 
$\CSlice_{-3}^3$, such that $\hLapse,\Shift$ are small in $W^{1,\infty}$
norm, defined by $||f||_{W^{1,\infty}} = ||f||_{L^\infty} +
||Df||_{L^\infty}$. 
It is straightforward to check
\begin{cor}\label{cor:NXgk-small}
Let $\alpha > 0$ be such that the conclusion of Theorem
\ref{thm:energy-norm} holds. 
There is a constant $\delta > 0$ so that for 
$(g,k) \in \Smallset(\alpha)$ with $\Energy(g,k) < \delta$, it holds that 
$$
\max(\Lambda, 
||\hLapse||_{W^{1,\infty}} , || \Shift||_{W^{1,\infty}}, || g
||_{W^{1,\infty}} , || k ||_{L^{\infty}} ) < 1/\delta . 
$$
\end{cor}
\begin{lemma}\label{lem:direct-est}
Let $\alpha>0$ be small enough so that the conclusion of Theorem
\ref{thm:energy-norm} holds. Let $(g,k) \in \Smallset(\alpha)$ and
let $\Lapse,\Shift$ be the corresponding solutions of the defining equations
(\ref{eq:defineNX}). Then there is a constant $C$ such that 
\begin{subequations}\label{eq:direct-est}
\begin{align}
|| \tilde{\hk} ||_{L^{\infty}} & \leq C \Energy^{\half} , \label{eq:hk-est}\\
|| \tilde{\hLapse} ||_{L^{\infty}} &\leq C \Energy , \label{eq:hLapse-est} \\
|| \widetilde{\nabla \Lapse} ||_{L^{\infty}} &\leq C \Energy ,
\label{eq:nablaLapse-est} \\
|| \tilde{\hpi} ||_{L^{\infty}} 
&\leq C \Energy^{\half} . \label{eq:hpi-energy-est}
\end{align}
\end{subequations}
\mnote{\Lars changed to $\Energy^{1/2}$ in
(\ref{eq:hpi-energy-est}), check that this is consistent later}
\end{lemma}
\begin{proof}
The inequality (\ref{eq:hk-est}) follows from the definition of $\Smallset$
and Sobolev imbedding. 
The Lapse equation (\ref{eq:defineN}) implies by the maximum principle, 
$$
|\hLapse| \leq \frac{3}{\tau^2}
\frac{||\hk||^2_{L^{\infty}}}{||k||^2_{L^{\infty}}} , 
$$
which gives (\ref{eq:hLapse-est}) after rescaling. 
A standard elliptic estimate
gives (\ref{eq:nablaLapse-est}). Finally,
(\ref{eq:picomp}) together with the above estimates yield 
$|| \tilde{\hpi} ||_{L^{\infty}} \leq 
C ( \Energy^{\half} + \Energy )$, which after using the smallness assumption
and redefining $C$ gives 
$$
|| \tilde{\hpi} ||_{L^{\infty}} \leq C \Energy^{\half} .
$$
\end{proof}

\subsection{Differential inequalities for the Rescaled Bel--Robinson
energies} \label{sec:BRdiff}

In this section we will estimate the time
derivatives 
$\partial_{\logtime} \widetilde{\EBR}_i$
of the scale--free Bel--Robinson energies with respect to the logarithmic
time $\logtime$ defined in section \ref{sec:BRvac}. 
\begin{lemma}\label{lem:EBR-0-est}
Assume $(g,k) \in \Smallset(\alpha)$ for $\alpha$ sufficiently small so
that the conclusion of Theorem \ref{thm:energy-norm} holds. 
Then 
\begin{equation} \label{eq:EBR-0-est}
 \partial_{\logtime} \widetilde{\EBR}_0(\logtime,W)  \leq 
- (2 - 2C\Energy^{1/2})  \widetilde{\EBR}_0 ,
\end{equation}
\end{lemma}
\begin{proof}
Replacing all the
fields in the RHS of (\ref{eq:BRconsJ=0}) by their scale--free versions, noting
in particular that $\tilde \tau = -3$ and $t < 0$, we get 
\begin{align*}
\partial_\sigma \tilde \EBR_{0} 
&= -2 \widetilde{\EBR}_{0}
+  \tilde F_1
\end{align*}
where $\tilde F_1$ is the scalefree version of 
$$
F_1 =  - 9 \int_{M_\tau}\Lapse \BR_{abcd}
\hpi^{ab} T^{c} T^{d} d\mu_{M_\tau} + 3\tau \int_{M_{\tau}}
\hLapse \BR_{TTTT}
$$
The  maximum principle applied to the Lapse equation
(\ref{eq:defineN}) implies, after a rescaling, that 
$\tilde{\Lapse} \leq \third$.
This gives the estimate
$$
\tilde F_1 
\leq C ( || \tilde{\hpi} ||_{L^{\infty}} 
+ || \tilde{\hLapse} ||_{L^{\infty}} )
\widetilde{\EBR}_0(\tau,W) . 
$$
To finish the proof note that by Lemma \ref{lem:direct-est}, 
$$
|| \tilde{\hpi} ||_{L^{\infty}} 
+ || \tilde{\hLapse} ||_{L^{\infty}} \leq C  \Energy^{1/2} ,
$$
using the smallness assumption. We write the resulting inequality in the
form (\ref{eq:EBR-0-est}) for convenience. 
\end{proof}
\begin{remark}
The proof of Lemma \ref{lem:EBR-0-est} gives the inequality 
$$
 \partial_{\logtime} \widetilde{\EBR}_0  \leq 
- (2 - C(||\tilde{\hpi}||_{L^{\infty}} 
+ || \tilde{\hLapse}||_{L^{\infty}})) 
\widetilde{\EBR}_0  ,
$$
which is valid {\em without} the smallness assumption. 
\qed
\end{remark}

\begin{lemma}\label{lem:EBR-1-est}
Assume $(g,k) \in \Smallset(\alpha)$ for $\alpha$ suffiently small so
that the conclusion of Theorem \ref{thm:energy-norm} holds. 
Then 
\begin{equation} \label{eq:EBR-1-est}
\partial_{\logtime} \Energy  \leq - (2 - 2C \Energy^{\half}) \Energy .
\end{equation}
\end{lemma}
\begin{proof}
In view of the smallness condition, the inequality (\ref{eq:hLapse-est}) and
Lemma \ref{lem:EBR-0-est}, we only need to consider 
$\partial_{\logtime} \widetilde{\EBR}_1$.
We proceed as in the proof of Lemma \ref{lem:EBR-0-est}, using 
the scale--free version of
(\ref{eq:BRconsJi}) taking into account $\tilde \tau = -3$, we get 
\begin{align*}
\partial_{\logtime} \widetilde{\EBR}_1 
&= - 2 \widetilde{\EBR}_1 
+  \tilde F_2
\end{align*}
where $\tilde F_2$ is the scale--free version of 
\begin{multline*}
F_2 =  -6 \int_{M_{\tau}} N \GG_1(W) d\mu_{M_\tau}
- 9 \int_{M_\tau}\Lapse \BR(\anabla_T W)_{abcd}
\hpi^{ab} T^{c} T^{d} d\mu_{M_\tau} \\
 +5\tau \int_{M_{\tau}}
\hLapse \BR(\anabla_T W)_{TTTT}
\end{multline*}
We see that in order to estimate $\tilde F_2$, we need to
estimate $\GG_1(W)$, which is given in Lemma \ref{lem:cons-laws}. 
Taking into account the detailed structure of $\GG_1$, 
cf. Remark \ref{rem:struct} we get the estimate
\begin{multline*}
\int N \GG_1(W) \leq 
C \Big{[}  || \hpi ||_{\infty}
 ( 1 + || k ||_{\infty})(\EBR_0+ \EBR_1 )  \\
+ \int_{M_{\tau}} N |E(\anabla_T W)|( |E(W)|^2 + |B(W)|^2) d\mu_{M_\tau}
\Big{]} . 
\end{multline*}
By the Holder inequality, 
$$
\int |E(\anabla_T W)|(|E(W)|^2 + |B(W)|^2) \leq C \EBR_1^{1/2} 
(|| E(W) ||_{L^4}^2 + ||B(W)||_{L^4}^2)  .
$$
By the Sobolev inequality, we may bound the scalefree version of 
$$
|| E(W) ||_{L^4}^2 + ||B(W)||_{L^4}^2
$$
by $C\Energy$. We now have 
the estimate 
\begin{align} 
\int \tilde N \tilde{\GG}_1  &\leq 
C \left (  || \tilde{\hpi} ||_{\infty}
 ( 1 + || \tilde{k} ||_{\infty}) \Energy 
+ \Energy^{3/2} \right ) \\
&\leq C \Energy^{3/2} ,
\end{align}
for $(g,k) \in \Smallset(\alpha)$. Proceeding similarly with the other terms
in $\tilde F_2$ yelds an inequality which we write in the form 
(\ref{eq:EBR-1-est}) for convenience. 
\end{proof}

\section{Global existence\label{sec:global}} 
Fix $\tau_0 < 0$ and let $(g(\tau_0),k(\tau_0))$ be data for Einstein's 
equations with mean 
curvature $\tau_0$ and assume that $(g(\tau_0),k(\tau_0)) \in 
\Smallset(\alpha)$ for an 
$\alpha > 0$ small enough so that the conclusion of Theorem
\ref{thm:energy-norm} holds. 

We have seen above that for small data, 
the second order scale--free Bel--Robinson energy 
$\Energy$ satisfies the differential inequality
(\ref{eq:EBR-1-est}).
We will use this to prove
\begin{thm}[Global existence for small data]\label{thm:global} Assume that
$M$ is rigid. 
Let $\alpha > 0$ be such that
the conclusion of Theorem \ref{thm:energy-norm} holds.
There is an $\epsilon \in (0,\alpha)$ small enough that if $(g^0,k^0) \in
\Smallset(\epsilon)$, then the maximal existence interval in mean curvature
time $\tau$, for the vacuum 
Einstein equations in CMCSH gauge, 
with data $(g^0, k^0)$ is of the form $(T_-,0)$. In particular, the CMCSH
vacuum Einstein equations have global
existence in the expanding direction for initial data in
$\Smallset(\epsilon)$. 

Here $\epsilon$ can be chosen as 
$$
\epsilon = D(\alpha)^{-1} \min(\delta,C^{-2}) ,
$$
where $D(\alpha)$ is defined in Theorem \ref{thm:energy-norm}, 
$\delta > 0$ is given by Corollary \ref{cor:NXgk-small},
and $C$ is the constant in (\ref{eq:EBR-1-est}). 
\end{thm}
\begin{proof}
Under the assumptions of the theorem,  by 
Theorem \ref{thm:energy-norm}, 
\begin{equation}\label{eq:smallenergy}
\Energy(g^0,k^0) < \min(\delta, C^{-2}) 
\end{equation}
holds.
Thus the conclusion of Corollary
\ref{cor:NXgk-small} holds. By the definition of
$\Smallset(\alpha)$ we may apply Theorem \ref{thm:AMlocal} to conclude that 
there is
a nontrivial maximal existence interval $(T_-, T_+)$ in mean curvature time
$\tau$, 
with $T_+ \leq 0$,
for $(g^0,k^0)$ in 
$H^3\times H^2$. We will assume $T_+ < 0$ and prove that
this leads to a contradiction, using energy estimates and the continuation
principle. 

Let $y(\logtime)$ be the solution to the initial value problem 
\begin{equation}\label{eq:model-eq}
\frac{dy}{d\logtime} = - 2 y + 2C y^{3/2} , 
y(\logtime_0) = y_0 .
\end{equation}
Then if $y_0 = \Energy(g^0,k^0)$, 
is such that $y(\logtime) < \infty$ for $\logtime \in [\logtime_0 ,
\logtime_+]$, we have 
$$
\Energy(\logtime) \leq y(\logtime), \quad \logtime \in [\logtime_0, \logtime_+) .
$$
The solution to (\ref{eq:model-eq}) is 
$$
y^{-1/2} = C + e^{\logtime - \logtime_0} (y_0^{-1/2} - C) ,
$$
if $y_0< C^{-2}$, and in this case $y(\logtime)< y(\logtime_0)$ for 
$\logtime \in (\logtime_0,\infty)$. 
This means that if (\ref{eq:smallenergy}) holds at $\logtime = \logtime_0$, 
it holds for 
$\logtime \in [\logtime_0, \logtime_+)$. 
By Theorem \ref{thm:energy-norm}, this implies that
$||\tilde g - \gamma||_{H^3} + ||\tilde k + \gamma||_{H^2}$ is uniformly 
bounded for 
$\logtime \in [\logtime_0, \logtime_+)$. 
By Corollary \ref{cor:NXgk-small},
this implies that the inequality
$$
\sup_{\logtime \in [\logtime_0 , \logtime_+)} 
(\Lambda[\ame] + ||D\ame||_{L^{\infty}} + || k ||_{L^{\infty}} )
< \delta^{-1} ,
$$
holds.

In view of the continuation
principle, Point \ref{point:cont} of Theorem \ref{thm:AMlocal}, 
this contradicts the assumption that
$(T_-, T_+)$ is the maximal existence interval in mean curvature time $\tau$, 
with $T_+ < 0$. It follows that $T_+ = 0$ which
completes the proof. 
\end{proof}

\subsection{Geodesic completeness\label{sec:complete}}
\begin{thm}\label{lem:complete}
Let $(M,g^0 , k^0)$ and $(\aM, \ame)$ be as in Theorem
\ref{thm:global}. 
Then $(\aM, \ame)$ is causally geodesically
complete in the expanding direction. 
\end{thm} 
\begin{proof}
By Theorem \ref{thm:global}, $(\aM, \ame)$ is globally foliated by CMC
hypersurfaces to the future of $(M, g^0 ,k^0)$, i.e. in the expanding
direction, with $t = \tr k \nearrow t_* =  0$. 

Let $c(\lambda)$ be a future directed causal geodesic, 
with affine parameter $\lambda$. 
Let 
$$
u = \frac{dc}{d\lambda} , \quad \la u , u \ra = \left \{ \begin{array}{l}
-1 \\ 0 \end{array} \right .  , 
$$
be the normalized velocity, where $\la \cdot , \cdot
\ra  = \ame( \cdot , \cdot)$. The geodesic equation is 
\begin{equation}\label{eq:lambdageod}
\anabla_u u = 0 .
\end{equation}
As $c$ is causal, we may use $t$ as parameter. Let 
$$
u^0 = dt(u) = \frac{dt}{d\lambda} . 
$$
In order 
to prove geodesic completeness in the expanding direction, it is
sufficient to prove that the solution to the geodesic equation exists for an
infinite interval of the affine parameter, i.e. 
$ \lim_{t \nearrow t_*} \lambda(t) = \infty $ or 
$$
\lim_{t \nearrow t_*} \int_{t_0}^t \frac{d\lambda}{dt} dt = \infty ,
$$
or using the definition of $u^0$, 
\begin{equation}\label{eq:diverg}
\lim_{t \nearrow t_*} \int_{t_0}^t \frac{1}{u^0} dt = \infty .
\end{equation}
Suppose that we are able to prove that $\Lapse u^0$ is bounded from above as $t
\nearrow t_*$. Then (\ref{eq:diverg}) holds precisely when 
$$
\lim_{t\nearrow t_*} \int_{t_0}^{t} \Lapse  dt = \infty .
$$
For a function $f$ on $\aM$, we have 
\begin{equation}\label{eq:dtf}
\frac{df(c(t))}{dt} = \frac{d\lambda}{dt} \frac{df(c(t))}{d\lambda}
= \left ( \frac{dt}{d\lambda} \right )^{-1} \frac{df(c(t))}{d\lambda} 
= \frac{1}{u^0} \anabla_u f .
\end{equation}
A calculation in local coordinates using the 3+1 form of $\ame$, gives
$T_{\mu} = - \Lapse  \delta_{\mu}^0$ where $\delta_{\mu}^{\nu}$ is the Kronecker 
delta, i.e. $\la T , V \ra = -\Lapse dt(V)$ for any $V$. This shows that 
$$
- \Lapse  u^0 = \la u,T\ra ,
$$
or 
\begin{equation}\label{eq:usplit}
u = \Lapse u^0 T + Y ,
\end{equation}
where $Y$ is tangent to $M$. 
Let $\epsilon = 0,1$. Then by assumption, $\la u,u\ra = - |\epsilon|$ which
using (\ref{eq:usplit}) gives
$$
 |Y|_g^2 = \Lapse^2 (u^0)^2 - |\epsilon| .
$$
In particular, we get the inequality
\begin{equation}\label{eq:Ybound}
|Y|_g \leq \Lapse u^0 .
\end{equation}
A computation in a Fermi propagated frame gives using (\ref{eq:usplit}),
\begin{equation}\label{eq:nablauT}
\anabla_u T 
= \Lapse u^0 \Lapse^{-1} \nabla_i \Lapse e_i - k_{ij} Y^j e_i  .
\end{equation}
By our choice of time orientation we have $\Lapse > 0$ and $u^0 > 0$ and $\tr k <
0$. 

We now compute using $\anabla_u u = 0$, (\ref{eq:usplit}) and 
(\ref{eq:nablauT})
\begin{align*}
\frac{d}{dt} \ln(\Lapse u^0) &= - \frac{1}{\Lapse u^0} \frac{d}{dt} \la u, T\ra \\
&= - \frac{1}{\Lapse(u^0)^2}  \la u , \anabla_u T \ra 
\\
&= - \frac{1}{\Lapse (u^0)^2} ( u^0 \nabla_Y \Lapse - k_{ij} Y^i Y^j )  \\
&= - \frac{\nabla_Y \Lapse}{\Lapse u^0} + \Lapse \hk_{ij} \frac{Y^i}{\Lapse
u^0} \frac{Y^j}{\Lapse u^0}
+ \Lapse  \frac{\tr k}{3} \frac{|Y|_g^2}{\Lapse^2 (u^0)^2} \\
\intertext{use (\ref{eq:Ybound}) and $\tr k < 0$,}
&\leq ||\nabla \Lapse ||_{L^{\infty};g} + ||N\hk||_{L^{\infty};g} \\
\intertext{use scaling properties of $N,\hk,g$, cf. section \ref{sec:BRvac},}
&\leq ||\widetilde{\nabla \Lapse}||_{L^\infty;\tg} + \lambda^{-1} || \tLapse
\tilde{\hk} ||_{L^\infty;\tg} \, ,
\end{align*}
with $\lambda = \tr k /3 = t/3$. 
By the proof of the global existence result Theorem \ref{thm:global}, we
have $\Energy(t) \leq C t^2$ and $\tg$ is close to $\gamma$. 
Therefore by Sobolev imbedding, we can relate the norms w.r.t. $\tg$ to
the norms w.r.t. $\gamma$ and we get 
$$
\frac{d}{dt} \ln(\Lapse u^0) \leq C \left ( || \widetilde{\nabla \Lapse}  ||_{L^{\infty}} 
+ \lambda^{-1} ||\tLapse \tilde{\hk} ||_{L^{\infty}} \right ) .
$$
Now an application of the estimate (\ref{eq:direct-est}) together with the decay of $\Energy$ gives 
\begin{align*}
|| \widetilde{\nabla \Lapse}  ||_{L^{\infty}} &\leq C t^2  , \\
|| \tLapse  \tilde{\hk} ||_{L^{\infty}} &\leq Ct , 
\end{align*}
which in view of $\lambda^{-1} t = 3$ gives 
$$
\ln(\Lapse u^0)(t) - \ln(\Lapse u^0)(t_0) \leq C ,
$$ 
and hence $\ln(\Lapse u^0) \leq C$ for some constant $C$ as $t \nearrow t_*
= 0$.

We have now proved that $Nu^0$ is bounded from above and therefore 
it is sufficient to prove that 
\begin{equation}\label{eq:Ndiverg}
\lim_{t\nearrow t_*} \int_{t_0}^{t} \Lapse  dt = \infty .
\end{equation}
Write $\Lapse  = \hLapse + \frac{3}{t^2}$ as in (\ref{eq:hLapsedef}). Using
(\ref{eq:direct-est}) and the scaling rule (\ref{eq:scaling}) to estimate
$\hLapse$ gives $\Lapse \geq C/t^2$ as $t \nearrow t_*=0$. This shows that  
(\ref{eq:Ndiverg}) holds and completes the proof of Lemma
\ref{lem:complete}.
\end{proof}

\noindent{\bf Acknowledgements:}
The authors are grateful to the Erwin Schr\"odinger Institute in Vienna,
l'Institute des Hautes \'Etudes Scientifiques in Bures-sur-Yvette, the Albert
Einstein Institute in Golm, the Universit\'e Paris VI and the Institute of
Theoretical Physics in Santa Barbara for hospitality and support while most
of this work was being carried out. 

\appendix
\section{Basic definitions and identities}\label{app:basic}
\mnote{think about how much of this to keep, also what indices you want to
use here etc} 
\subsection{Conventions}\label{sec:basic}
We begin by recalling some basic facts and definitions. We use the following 
conventions for curvature.

The Riemann tensor is defined by 
$$
R(X,Y)Z = \nabla_X \nabla_Y Z - \nabla_Y \nabla_X Z - \nabla_{[X,Y]}Z .
$$
In a coordinate frame $\{e_a\}$ we have 
$$
R^d_{\ cab}Z^c = \nabla_a \nabla_b Z^d - \nabla_b \nabla_a Z^d .
$$
This gives the conventions for index calculations 
\begin{align} 
[\nabla_a, \nabla_b]t_c &= R_{abc}^{\ \ \ d} t_d , \\
[\nabla_a, \nabla_b]t^c &= R^c_{\ dab} t^d .
\end{align}

The Ricci curvature and the scalar curvature are defined (in an ON frame)
by 
\begin{align} 
\Ric(X,Y) &= \sum_i \la R(e_i,X)Y,e_i \ra  , \\
\Scal &= \sum_i \Ric(e_i, e_i) ,
\end{align}
or in index notation
$$
R_{ij} = g^{kl} R_{ikjl}, \qquad R = g^{ij} R_{ij} .
$$
Note also 
$$
\Ric(X,Y) = \tr (Z \mapsto R(Z,X)Y) .
$$
The Riemann tensor satisfies the Bianchi identities
$$
\nabla_{[e} R_{ab]cd} = \third ( \nabla_e R_{abcd} + \nabla_{a} R_{becd} 
+ \nabla_b R_{eacd} ) = 0 .
$$
The trace free part of the Riemann tensor in an $n$--dimensional manifold 
is 
\begin{equation}\label{eq:weyldef}
\begin{split}
C_{abcd} &= R_{abcd} - \frac{1}{n-2} ( g_{ac} R_{bd} + g_{bd} R_{ac} - g_{bc}R_{ad}
- g_{ad} R_{bc} )  \\
&\quad + \frac{1}{(n-1)(n-2)} ( g_{ac} g_{bd} - g_{ad} g_{bc} ) R .
\end{split} 
\end{equation}
The totally anti-symmetric tensor $\Eps$ in dimension $3+1$ 
satisfies the identities
\begin{subequations}
\begin{equation}
\begin{aligned}
\Eps^{\alpha_1\alpha_2\alpha_3\alpha_4}\Eps_{\beta_1\beta_2\beta_3\beta_4}
&= -\det(\delta^{\alpha_i}_{\beta_j})_{i,j=1,\dots,4} ,  \\
\Eps^{\alpha_1\alpha_2\alpha_3\alpha_4}\Eps_{\alpha_1\beta_2\beta_3\beta_4}
&= -\det(\delta^{\alpha_i}_{\beta_j})_{i,j=2,\dots,4} ,
\\
\Eps^{\alpha_1\alpha_2\alpha_3\alpha_4}\Eps_{\alpha_1\alpha_2\beta_3\beta_4}
&= -2\det(\delta^{\alpha_i}_{\beta_j})_{i,j=3,\dots,4} ,  \\
\Eps^{\alpha_1\alpha_2\alpha_3\alpha_4}\Eps_{\alpha_1\alpha_2\alpha_3\beta_4}
&= -6\delta^{\alpha_4}_{\beta_4}  , \\
\Eps^{\alpha_1\alpha_2\alpha_3\alpha_4}\Eps_{\alpha_1\alpha_2\alpha_3\alpha_4}
&= -24 .
\end{aligned}
\end{equation}
\end{subequations}
In an ON frame adapted to a spacelike hypersurface $M$ in a $3+1$ dimensional manifold
we define 
(c.f. \cite[p. 144]{christo:klain})
\begin{equation}
\Eps_{ijk} = \Eps_{T ijk} .
\end{equation} 
\begin{subequations}
\begin{align}
\Eps^{i_1 i_2 i_3} \Eps_{j_1 j_2 j_3} &=
\det(\delta^{i_k}_{j_l})_{k,l=1,2,3}  
&= 6 \delta^{[i_1}_{j_1} \delta^{i_2}_{j_2} \delta^{i_3]}_{j_3} ,
\\
\Eps^{i_1 i_2 i_3} \Eps_{i_1 j_2 j_3} &= \det( \delta^{i_k}_{j_l})_{k,l=2,3})
&= 2 \delta^{[i_2}_{j_2} \delta^{i_3]}_{j_3} ,
\\
\Eps^{i_1 i_2 i_3} \Eps_{i_1 i_2 j_3} &= 2 \delta^{i_3}_{j_3} ,
\\
\Eps^{i_1 i_2 i_3} \Eps_{i_1 i_2 i_3} &= 6 .
\end{align}
\end{subequations}
In dimension $3$ we have the duality relations
$$
\xi_{ab} = \Eps_{ab}^{\ \ m} \eta_m ,
$$
for $\xi_{ab} = \xi_{[ab]}$, where 
$$
\eta_m = \half \Eps_m^{\ ab} \xi_{ab} .
$$

\subsection{Operations on symmetric 2-tensors}\label{sec:operations}
Define the following operations on symmetric 2-tensors on a 3--dimensional
Riemann manifold:
\begin{align}
A \cdot B &= A_{ab} B^{ab} , \label{eq:AdotB} \\
(A \wedge B)_a &= \Eps_a^{\ bc} A_b^{\ d}B_{dc} , 
\label{eq:symsymwedge} \\
(v \wedge A)_{ab} &= \Eps_a^{\ cd} v_c A_{db} + \Eps_b^{\ cd}v_c A_{ad} , 
\label{eq:vecsymwedge} \\
(A \times B)_{ab} &= \Eps_a^{\ cd} \Eps_b^{\ ef} A_{ce} B_{df} + \third (A
\cdot B) g_{ab} - \third (\tr A)(\tr B) g_{ab}  , 
\label{eq:symsymcross} \\
\curl A_{ab} &= \half ( \Eps_a^{\ cd} \nabla_d A_{cb} + \Eps_b^{\ cd}
\nabla_d A_{ca}) , 
\label{eq:symcurl} \\
\div A_a &= \nabla^b A_{ab} . 
\label{eq:symdiv}
\end{align}
The operation $\wedge$ is skew symmetric, while $\times$ is symmetric, and
the identities
\begin{align*}
A \cdot (v \wedge B) &= - 2 v \cdot (A \wedge B) \\
A \cdot (B \times C) &= (A \times B) \cdot C &\text{(if $\tr A = \tr C = 0)$}
\end{align*}
hold. 
The expression $A\times B$ can be expanded as 
\begin{multline*}
(A \times B)_{ab} = A_a^{\ c}B_{cb} +  A_b^{\ c} B_{ca} \\
- \frac{2}{3} ( A \cdot B) g_{ab} 
+ \frac{2}{3} (\tr A)(\tr B) g_{ab}  - (\tr A) B_{ab} -(\tr B)A_{ab} .
\end{multline*}
A computation shows 
\begin{equation}\label{eq:divAwedgeB}
\div (A \wedge B) = -(\curl A) \cdot B + A \cdot (\curl B) .
\end{equation}

Let $A$ be a symmetric covariant 2-tensor on $\aM$ and suppose $A$ is
$t$--tangent, i.e. $A_{\alpha\beta} T^\beta = 0$. 
Then in a Fermi propagated frame, 
\begin{align}
\anabla_T A_{ij} &= T A_{ij} \\
\Lie_{\partial_t} A_{ij} &= \Lapse \anabla_T A_{ij} 
- \Lapse \left ( (k \times A)_{ij} + \frac{2}{3} (k \cdot A) g_{ij}
\right. \nonumber \\
& \quad \left. 
- \frac{2}{3} (\tr k) (\tr A) g_{ij} 
+ (\tr A)k_{ij} + (\tr k) A_{ij} \right ) .  \label{eq:Liedt}
\end{align}    
Define the covariant exterior derivative  $d^{\nabla}u$ on symmetric
2-tensors by 
\begin{equation}\label{eq:extderdef}
(d^{\nabla} u)_{ijk} = \nabla_k u_{ij} - \nabla_j u_{ik}  .
\end{equation}
The operators $\curl, \div, d^{\nabla}$ are related by 
\cite[p. 103]{christo:klain}
\begin{equation}\label{eq:dnabla-curl-div}
d^{\nabla} u_{kij} = \left (\curl u_{kl} + \half (\div u_m - \nabla_m \tr u) 
\Eps^m_{\ kl} \right )\Eps^l_{\ ij} .
\end{equation}
Taking into account the symmetry of $\curl$ this implies 
\begin{equation}\label{eq:norm-dnabla}
|d^{\nabla} u|^2 = 2 (|\curl u|^2 + \half |\div u - \nabla \tr u|^2) .
\end{equation}

If $u$ has compact support then in dimension $n$, 
\begin{equation}\label{eq:dnabla-id}
\int_M | d^{\nabla} u|^2 =   2 \int_M  |\nabla u|^2 - 2 \int_M |\div u |^2 + 
2 \int_M ( u_k^{\ l} R^k_{\ ijl} + u_{ik} R^k_{\ j}) u^{ij} .
\end{equation}
This leads to, if $\tr u = 0$,  
\begin{equation}\label{eq:hodge-id}
\int_M ( | \nabla u|^2 + 3 R_{ij} u^{ki} u_k^{\ j} - \half R |u|^2)  = 
\int_M ( |\curl u|^2 + \frac{3}{2} |\div u|^2) ,
\end{equation}
in case $M$ is of dimension 3. If we further restrict to $(M,\gamma)$ with
$\gamma$ hyperbolic, so that $R[\gamma] = -6$, we get 
\begin{equation}\label{eq:quadeq}
\int_M ( |\curl u|^2 + \frac{3}{2} |\div u|^2) = 
\int_M (|\nabla u|^2 - 3 |u|^2)  .
\end{equation}


\providecommand{\bysame}{\leavevmode\hbox to3em{\hrulefill}\thinspace}
\providecommand{\MR}{\relax\ifhmode\unskip\space\fi MR }
\providecommand{\MRhref}[2]{%
  \href{http://www.ams.org/mathscinet-getitem?mr=#1}{#2}
}
\providecommand{\href}[2]{#2}

\end{document}